\documentclass[twocolumn,epjc3]{svjour3}
\input{Preamble.tex}
\begin{document}

\title{Sensitivity of the Hyper-Kamiokande experiment to neutrino oscillation parameters using accelerator neutrinos}

% The author list
% ----- Start automatically generated HyperK info
% ----- Start author list
\renewcommand{\thefootnote}{\alph{footnote}}
\author{
K.~Abe\thanksref{a} 
\and
R.~Ahl~Laamara\thanksref{b} 
\and
H.~Aihara\thanksref{c} 
\and
A.~Ajmi\thanksref{d} 
\and
R.~Akutsu\thanksref{e,alsoa} 
\and
H.~Alarakia-Charles\thanksref{f} 
\and
I.~Alekseev\thanksref{} \orcidlink{0000-0003-3358-9635}
\and
Y.~Alj~Hakim\thanksref{g} \orcidlink{0000-0002-9065-1303}
\and
S. Alonso Monsalve\thanksref{h} \orcidlink{0000-0002-9678-7121}
\and
E.~Amato\thanksref{i}
% ...
}
% ----- End author list
% ----- Start affiliation list
\institute{
\label{a}Kamioka Observatory, Institute for Cosmic Ray Research, University of Tokyo, Kamioka, Gifu, 506-1205 Japan
\and
\label{b}Faculty of Sciences, Mohammed V University, Rabat
\and
\label{c}University of Tokyo, Department of Physics, 7-3-1 Hongo, Bunkyo-ku, Tokyo, 113-0033, Japan
\and
\label{d}University of Winnipeg, 515 Portage Ave, Winnipeg, Manitoba, R3B 2E9, Canada
\and
\label{e}High Energy Accelerator Research Organization (KEK), 1-1 Oho, Tsukuba, Ibaraki, 305-0801, Japan
\and
\label{f}Physics Department, Lancaster University, Lancaster, LA1 4YB, United Kingdom
\and
\label{g}University of Sheffield, Department of Mathematical and Physical Sciences, Western Bank, Sheffield, S10 2TN, United Kingdom
\and
\label{h}ETH Zurich, Institute for Particle Physics and Astrophysics, Otto-Stern-Weg 5, Zurich, CH-8093, Switzerland
\and
\label{i}INFN Sezione di Bari, via Orabona 4, Bari, 70126, Italy
% ...
\and
\label{alsoa}also at J-PARC, Ibaraki, Japan
% ...
}
% ----- End affiliation list
% ----- End automatically generated HyperK info

\date{Received:  / Accepted: }
\onecolumn

\maketitle
\twocolumn

\begin{abstract}
This paper presents the expected sensitivity to the neutrino oscillation parameters of the Hyper-Kamiokande long-baseline program. The Hyper-Kamiokande experiment, currently under construction in Japan, will measure the oscillations of accelerator-produced neutrinos with thousands of selected events per sample: this corresponds to an increase of statistics
of a factor 25 to 100 with respect to recent results from the currently-running long-baseline neutrino oscillation experiment in Japan, T2K. 
In the most favorable scenario we will achieve the discovery of Charge-Parity (CP) violation in neutrino oscillation at $5\sigma$ C.L. in less than three years. With 10 years of data-taking, and assuming a neutrino : antineutrino beam running ratio of 1:3, a CP violation discovery at $5\sigma$ C.L. is possible for more than 60\% of the actual values of the CP-violating phase, $\delta_{CP}$.  
Moreover, we will measure $\delta_{CP}$ with a precision ranging from 20\(\degree\), in the case of maximal CP violation, to 6\(\degree\), in the case of CP conservation. We aim to achieve a 0.5\% resolution on the $\Delta m^2_{32}$ parameter, and a resolution between 3\% and 0.5\% on the $\sin^2\theta_{23}$ parameter, depending on its true value. These results are obtained by extending the analysis methods of T2K with dedicated tuning to take into account the Hyper-Kamiokande design: the larger far detector, the more powerful beam, the upgraded near detector ND280, and the planned additional Intermediate Water Cherenkov Detector.
\end{abstract}
%\twocolumn

\section{Introduction}
\label{sec:introduction}

Neutrinos propagate as mass eigenstates ($\nu_1$, $\nu_2$, $\nu_3$), while they interact with matter as flavour eigenstates of the weak interaction ($\nu_e$, $\nu_\mu$, $\nu_\tau$), coupling to the corresponding charged leptons (electron, muon and tau).
In the standard paradigm of neutrino oscillations, the three mass eigenstates can be expressed as an admixture of the flavour eigenstates using a $3\times3$ unitary mixing matrix ($U_{\alpha k}; \alpha=e,\mu,\tau; k=1,2,3$), so called Pontecorvo-Maki-Nakagawa-Sakata (PMNS)~\cite{Pontecorvo:1967fh,Maki:1962mu} matrix. Out of the nine degrees of freedom of such a matrix, five could be reabsorbed as unphysical phases in the definition of the lepton fields. In the most used parametrization~\cite{ParticleDataGroup:2024cfk} the remaining degrees of freedom are encoded in a SO(3) matrix, using Tait-Bryan rotation angles ($\theta_{12}$,~$\theta_{13}$,~and $\theta_{23}$), and an additional complex phase $\delta_{CP}$, 
which parameterizes a possible Charge-Parity violation (CPV) in the lepton sector.
Considering the time evolution of the mass eigenstates during their propagation, the neutrino oscillation probability also depends on the squared mass difference between the pairs of mass eigenstates ($\Delta m^2_{ij}=m^2_i-m^2_j$). 

In this paper we present the sensitivity of the Hyper-Kamiokande experiment to measuring the oscillation parameters using accelerator neutrinos. 
As of today, all three mixing angles, as well as the $\Delta m^2_{21}$ and $|\Delta m^2_{32}|$ parameters have been measured with a few percent precision or better~\cite{ParticleDataGroup:2024cfk}. The phase $\delta_{CP}$ is still unknown, first hints from T2K~\cite{T2K:2023smv} point to large CPV, but they are not confirmed by NOvA~\cite{NOvA:2023iam}. A possible definitive discovery of this new fundamental source of CPV, the first in the lepton sector, would have profound implications on the comprehension of the matter-antimatter asymmetry in the Universe, in particular in the framework of leptogenesis~\cite{Fukugita:1986hr} with low-energy seesaw mechanisms~\cite{Granelli:2023tcj}.  CPV discovery is the primary target of the Hyper-Kamiokande sensitivity analysis reported in this paper.  The sign of $\Delta m^2_{32}$, also known as the mass ordering, is still unknown, with first indications from Super-Kamiokande~\cite{Super-Kamiokande:2023ahc} atmospheric neutrino measurements showing a preference for normal ordering ($m_3 > m_2$). The octant of the $\theta_{23}$ mixing angle is still unknown, leaving the possibility of maximal mixing between muon and tau neutrinos ($\theta_{23}=\pi/4$) open. 
In the next decade these neutrino oscillation parameters will be also measured by another long-baseline experiment, DUNE~\cite{DUNE:2021mtg}, in construction in US.

From a broader perspective, the present oscillation paradigm consists of an effective parametrization of flavour mixing: whether or not the specific values of the oscillation parameters are due to an underlying fundamental symmetry or an underlying basic principle remains an open question. 
Precise measurements of these parameters, within the scope of the present Hyper-Kamiokande analysis, may hint at, or at least help to discard, specific flavour symmetry models~\cite{Gehrlein:2022nss}.

\subsection{The Hyper-Kamiokande experiment}
Hyper-Kamiokande~\cite{Hyper-KamiokandeProto-:2015xww,Hyper-Kamiokande:2018ofw} is an experiment under construction aiming to perform precision measurements of 
neutrino oscillations and determine whether neutrinos violate Nature's Charge-Parity (CP) symmetry. It will also perform the world's most sensitive search for proton decay, supernova neutrinos detection and other physics measurements.
Hyper-Kamiokande is the third-generation water Cherenkov neutrino detector in Japan, following Kamiokande~\cite{Kamiokande-II:1987idp} and Super-Kamiokande~\cite{Super-Kamiokande:2002weg}, the experiment currently underway. Hyper-Kamiokande will measure oscillations of atmospheric and solar neutrinos as well as oscillations of neutrinos produced by an accelerator, as in its predecessor long-baseline experiments K2K (KEK to Kamioka)~\cite{K2K:2006yov} and T2K (Tokai to Kamioka)~\cite{T2K:2011qtm}.

The Hyper-Kamiokande experiment consists of the existing J-PARC neutrino beam and a set of near detectors, both currently used for the T2K experiment and being upgraded to increase performance, and two new water Cherenkov detectors, an intermediate detector of about 600-ton at around 1~km from the neutrino beam production target and a 258,000-ton far detector at 295~km (oscillation baseline).

The J-PARC accelerator complex provides a beam of 30~GeV protons and is being upgraded to reach 1.3~MW power near the beginning of the Hyper-Kamiokande data taking~\cite{T2K:2019eao}. The proton beam impinges on a graphite target, producing hadrons (primarily pions and kaons) which are focused and charge-selected by three electromagnetic horns. The hadrons are thus guided to a 96~m long He-filled vessel, where they decay generating a flux of charged leptons and neutrinos. The produced neutrino flux is highly dominated by muon neutrinos, with a small background ($<$1\%) of electron neutrinos. A flux dominated by neutrinos (Forward Horn Current, FHC) or antineutrinos (Reverse Horn Current, RHC) can be produced by inverting the polarity of the horns. Hyper-Kamiokande will rely on the off-axis technique, which profits from the kinematic properties of the two-body pion decay: by placing the near and far detectors at 2.5\(\degree\) off of the beam axis, the neutrino energy is peaked at 600~MeV, corresponding to maximal neutrino oscillations at a propagation baseline of 295~km, while the fraction of beam electron neutrinos, mostly coming from muon and kaon decays, is reduced.

The Hyper-Kamiokande far detector will be built 295~km from the production target and situated beneath the peak of Mt. Nijyugo, resulting in a 1,750 meters-water-equivalent overburden.  The detector is a 71~m high and 68~m diameter cylinder filled with ultra-pure water and split into optically isolated inner and outer detector regions.  The outer detector will consist of a 1 (2)~m wide shell surrounding the inner detector barrel (top and bottom) and is planned to be instrumented with approximately 3,600 8~cm diameter photomultiplier tubes (PMTs), each equipped with a 30~cm$\times$30~cm wavelength shifting plate to enhance light collection. The outer detector will identify and veto entering charged particles, such as cosmic muons or particles created by neutrino interactions in the surrounding rock.
The inner detector, 
acting as main target for the neutrino interactions, 
has a volume of about 217~kton and will be instrumented with approximately 20,000 50~cm diameter PMTs and 1,000 multi-PMT photosensor modules~\cite{Hyper-Kamiokande:2018ofw}.  The photo-cathode coverage of Hyper-Kamiokande will be about 20\%, compared to 40\% at Super-Kamiokande.
However, the light collection efficiency of the 50~cm PMTs is doubled.
The photon detection efficiency of Hyper-Kamiokande will therefore be approximately equal to that of Super-Kamiokande.

The set of near detectors placed 280~m from the beam target includes INGRID~\cite{Abe:2011xv}, located on-axis for beam position and direction monitoring, and ND280, a magnetized multipurpose detector which measures the neutrino flux and neutrino-nucleus interaction cross-sections. The ND280 detector has been recently upgraded~\cite{T2K:2019bbb} to increase the detectable kinematic range of particles from neutrino interactions, in view of the additional statistics still to be collected by the T2K experiment. The upgraded ND280 detector will also serve the Hyper-Kamiokande experiment.  ND280 consists of two main tracking regions: two vertical targets composed of sets of perpendicular scintillating bars (one target also hosting bags of passive water) interleaved with three vertical Time Projections Chambers; and a horizontal, highly granular, scintillator detector sandwiched between two horizontal Time Projection Chambers and further surrounded by scintillating panels for Time of Flight measurement. The detectors are surrounded by an electromagnetic calorimeter and embedded into a 0.2~T magnet, which also hosts a muon range detector.

Alongside the upgraded ND280, a new water Cherenkov detector will be built approximately 1~km from the neutrino production target~\cite{Hyper-Kamiokande:2018ofw}.  The Intermediate Water Cherenkov Detector (IWCD)  preliminary design consists of 
an 8.8~m diameter and 10~m tall cylinder with a 7~m diameter and 8~m tall inner detector region.  
The inner detector will be instrumented with about 350 multi-PMT modules and will have a target mass of approximately 300~tonnes.  The detector will be placed in a vertical pit allowing the detector to be positioned anywhere between $4$\(\degree\) and $1.5$\(\degree\) off the central axis of the neutrino beam. IWCD can directly measure the relationship between true and reconstructed neutrino energy by sampling different off-axis angles~\cite{nuPRISM:2014mzw}.  The large target mass, high precision electron/muon discrimination and self-shielding property of water also allow IWCD to make precise measurements of electron neutrino and antineutrino cross-sections.

\subsection{Neutrino Oscillation Probabilities}

The Hyper-Kamiokande accelerator neutrino flux is peaked at about 600~MeV, an energy where matter effects are subdominant ($<$10\%). While the oscillation formulas used in the analysis described here correctly include matter effects, oscillation formulas in vacuum will be described in this section as a simplified paradigm to show the sensitivity of the various Hyper-Kamiokande samples to the oscillation parameters. We can consider four main channels: muon (anti)neutrino disappearance and electron (anti)neutrino appearance. While most muon (anti)neutrinos oscillate into tau (anti)neutrinos, tau production from charged-current interactions is possible only in the high-energy tail of the Hyper-Kamiokande (anti)neutrino flux, thus far from the oscillation maximum. For this reason this sample is not considered in this paper. 

The muon (anti)neutrino disappearance formula in vacuum, relying on the PMNS parametrization of the mixing matrix, reads
\begin{multline}
P(\barparen{\nu_\mu}\rightarrow\barparen{\nu_\mu}) \simeq 1-4\cos^2\theta_{13}\sin^2\theta_{23} \\\times (1-\cos^2\theta_{13}\sin^2\theta_{23}) \sin^2\left(1.27\frac{\Delta m_{32}^2 L}{E}\right),
 \label{eq:Pdis}
\end{multline}
with $L$ as the oscillation  baseline, in km, $E$ as the neutrino energy, in GeV, and $\Delta m_{32}$ in eV.
The formula is independent on $\delta_{CP}$, thus applies to both neutrino and antineutrino disappearance in the standard oscillation paradigm. 
Exotic oscillation scenarios based on new physics models~\cite{Arguelles:2022tki}, such as CPT violation through comparisons of muon neutrino and antineutrino oscillation probabilities, are not covered in this study.
As can be seen from Eq.~\ref{eq:Pdis}, the muon (anti)neutrino samples are sensitive to the so-called atmospheric parameters, $|\Delta m^2_{32}|$ and $\sin^2\theta_{23}$. 
On the other hand, these samples cannot measure the sign of $\Delta m^2_{32}$, i.e. the mass ordering, and they suffer from a degeneracy of the $\theta_{23}$ octant,  i.e. cannot distinguish $[0,\pi/4]$ from $[\pi/4,\pi/2]$.
We can resolve this degeneracy using electron (anti)neutrino appearance samples. 
The formula for electron (anti)neutrino appearance in vacuum, in the $\Delta m^2_{21}/\Delta m^2_{31} \ll 1$ approximation, is
\begin{multline}
 P(\barparen{\nu_\mu} \rightarrow \barparen{\nu_e})\simeq  4c_{13}^2s_{13}^2s_{23}^2\cdot \sin^2\Delta _{31} \\
 + 8c_{13}^2s_{12}s_{13}s_{23}(c_{12}c_{23}\cos\delta_{CP}-s_{12}s_{13}s_{23}) \\ \qquad \qquad \qquad \qquad \cdot \cos\Delta_{32}\cdot \sin\Delta_{31}\cdot \sin\Delta_{21}\\
 - (+) 8c_{13}^2c_{12}c_{23}s_{12}s_{13}s_{23}\sin\delta_{CP}\cdot \sin\Delta_{32}\cdot \sin\Delta_{31}\cdot \sin\Delta_{21}\\
 + 4c_{13}^2s_{12}^2(c_{12}^2c_{23}^2+s_{12}^2s_{13}^2s_{23}^2-2c_{12}c_{23}s_{12}s_{13}s_{23}\cos\delta_{CP}) \\\cdot \sin^2\Delta_{21},
 \label{eq:Pap}
 \end{multline}
where we shortened $s_{ij}=\sin\theta_{ij}$,  $c_{ij}=\cos\theta_{ij}$, \linebreak $\Delta_{ij}=1.27 \Delta m^2_{ij} L/E$ (same units as Eq.~\ref{eq:Pdis}) and the term depending on $\sin\delta_{CP}$ is CP-odd, thus changing sign for antineutrinos. Hyper-Kamiokande will, therefore, feature a direct sensitivity to the CP asymmetry between neutrino and antineutrino oscillations, notably thanks to the capability of the beamline to produce a clean flux of neutrinos (in FHC) and antineutrinos (in RHC).
Inclusion of constraints from solar neutrino measurements~\cite{Davis:1968cp,GNO:2005bds,SAGE:2009eeu,Bellini:2011rx,SNO:2011hxd,Super-Kamiokande:2016yck} and KamLAND experiment~\cite{KamLAND:2013rgu} on the $\theta_{12}$ and $\Delta m^2_{21}$ parameters and from the reactor experiments on the $\theta_{13}$ angle~\cite{DayaBay:2018yms,RENO:2019otc,DoubleChooz:2019qbj} further enhance such sensitivity. 

When considering matter effects in the oscillation equations~\cite{Barger:1980tf,Prob3}, the (anti)neutrino appearance samples also feature a sensitivity to the mass ordering, manifesting itself as an additional asymmetry between electron neutrino and antineutrino appearance probabilities. At the energy of the Hyper-Kamiokande flux, matter effects are minor and they are degenerate with CPV effects, except in  the two extreme cases where matter effects push the neutrino/antineutrino asymmetry beyond what is allowed by maximal CPV (notably, $\delta_{CP}=-\pi/2$ and normal ordering or $\delta_{CP}=\pi/2$ and inverted ordering). In Hyper-Kamiokande, the mass ordering will be 
measured with atmospheric neutrinos: the combination of atmospheric and beam neutrinos~\cite{Hyper-Kamiokande:2018ofw} reaches between 3.5 and 4.5$\sigma$ MO determination in 6 years of data taking, depending on the value of $\sin^2\theta_{23}$ in the 1$\sigma$ interval from presently available oscillation measurements.
Indeed, mass ordering can also be measured by other experiments apart from long-baseline approaches: various running and forthcoming atmospheric and reactor experiments~\cite{Super-Kamiokande:2023ahc,JUNO:2024jaw,KM3NeT:2021ozk,IceCube:2023ins} 
feature high sensitivity to mass ordering.
In the following, we will therefore assume that the mass ordering is known and is normal. Whenever this assumption could have a sizable impact on the results, e.g. delaying the reach of CPV discovery, we will explicitly mention it.

\section{Event simulation}
\label{sec:events}
\subsection{Assumed exposure}
Hyper-Kamiokande is expected to collect statistics from $2.7\times10^{21}$ Protons-On-Target (POT) per calendar year, corresponding to 6 cycles of 22 days with 87$\%$ running efficiency at 1.3~MW. We chose a partition of $1/4$ of the exposure in FHC and $3/4$ in RHC, considering proper control of systematic uncertainties and direct access to extensive statistics of both neutrino and antineutrino events for a robust assessment of CP violation. In particular, this FHC/RHC partition is the optimal one to break the CPV degeneracy with $\sin^2\theta_{23}$ and $\sin^2\theta_{13}$. Due to difference in the cross-section and flux of neutrinos and antineutrinos, this partition leads to a comparable number of events in the FHC and RHC electron-like samples in the case of CP symmetry conservation.

\begin{figure*}[ht]
\centering
        \includegraphics[width=0.48\textwidth]{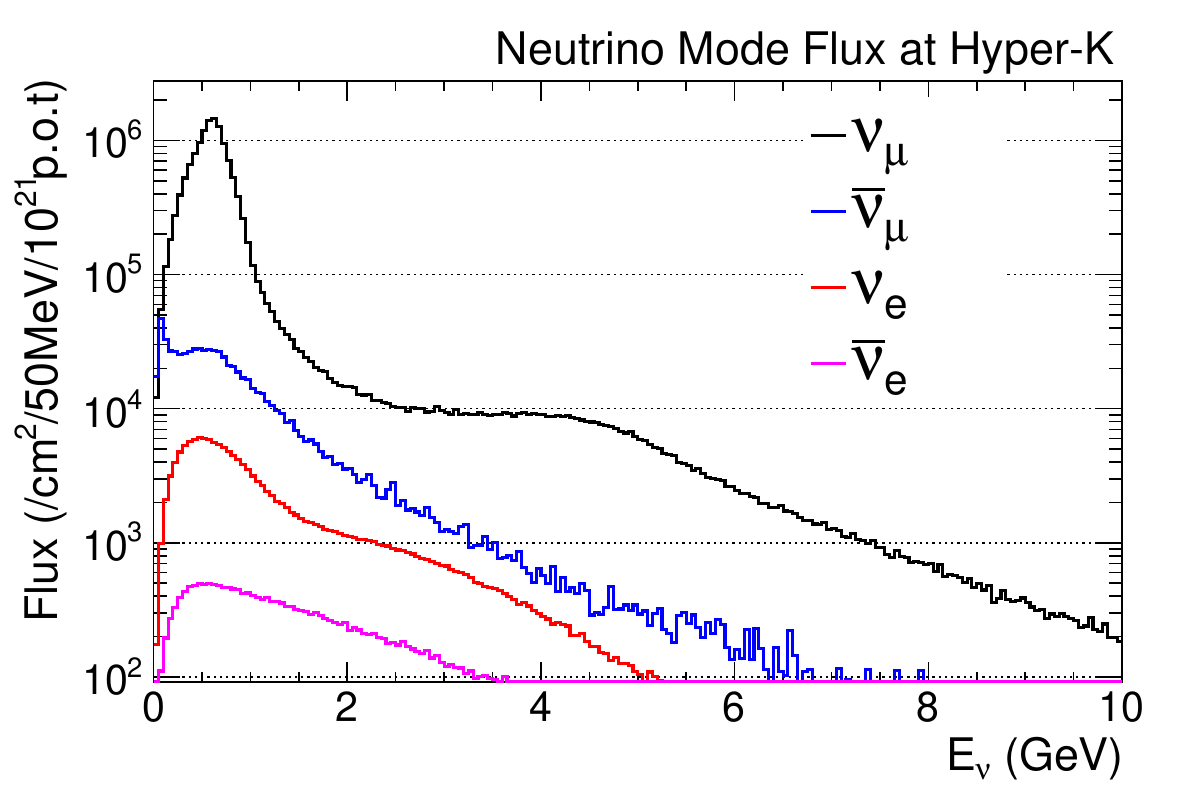}
        \includegraphics[width=0.48\textwidth]{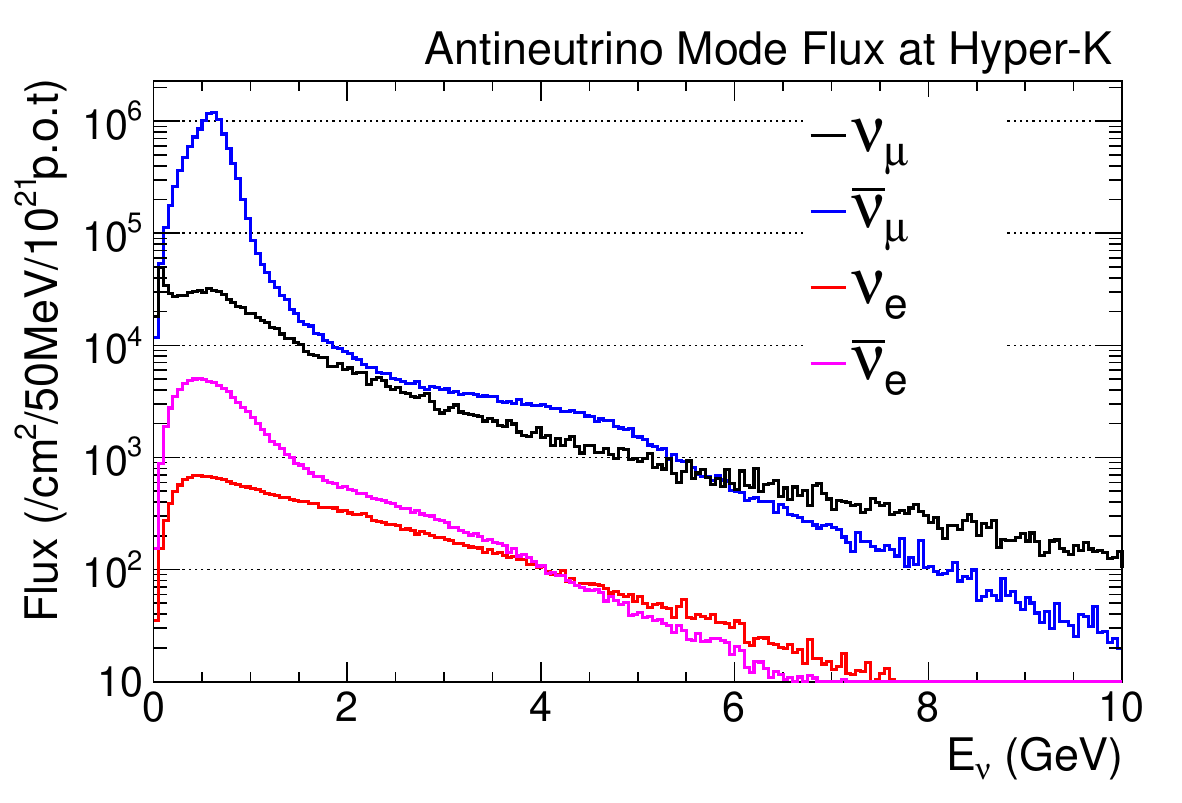}
   \caption{Simulated flux at the far detector in neutrino mode (left) and antineutrino mode (right).}
    \label{Flux}
\end{figure*}

We use the neutrino flux model of the T2K experiment~\cite{T2K:2012bge} as the basis for the analysis shown here. The expected neutrino flux has been weighted to take into account the expected increase in horn current (from $250$~kA used in the simulation of the T2K flux to $320$~kA for Hyper-Kamiokande) and the different relative position of the far detector (same off-axis angle but on the opposite side of the beam center). The simulated flux is shown in Fig.~\ref{Flux}. The initial flux uncertainties (i.e. without any constraint by the near detector) are the same as in Ref.~\cite{T2K:2023smv}, which profit from the NA61/SHINE hadroproduction measurements using the T2K replica target from Ref.~\cite{NA61SHINE:2016nlf}.

\subsection{Event samples}
To model the expected event rates and kinematics at the far detector, we use the T2K Monte Carlo simulation. Full details of the simulation, event reconstruction and event selection are described in Ref.~\cite{T2K:2023smv}. All events must be fully contained in the inner detector and to have only one prompt reconstructed particle, an outgoing lepton, in order to enhance the fraction of quasi-elastic events where the neutrino energy can be estimated more accurately. The events are then separated into electron-like (1Re) and muon-like (1R$\mu$) samples, and a sample-dependent fiducial volume cut is applied. Further cuts are applied to remove background events, such as pions produced by neutral current neutrino interactions.  Finally, the electron-like events are required to have a maximum reconstructed neutrino energy of $1.25$~GeV, since higher-energy beam neutrinos are insensitive to oscillations and prone to mismodeling of systematic uncertainties.  The samples are separated depending on the horn current, between neutrino- (FHC) or antineutrino-dominated (RHC) beam. 
We include an additional sample of electron-like events that have one delayed triggered signal relative to the primary interaction, consistent with a Michel electron from an unseen, positively-charged pion decay chain (1Re1De). This sample is only included for the FHC beam mode and allows to exploit also events with single pion production.

A scaling has been applied to the generated Monte-Carlo events to take into account the increased size of the Hyper-Kamiokande far detector compared to Super-Kamiokande.  Events are separated according to the distance from the interaction vertex to the tank wall in the lepton's direction of travel (so-called ``ToWall'' parameter).  Events with ToWall larger than $200$~cm are scaled by the ratio of the  detectors' fiducial volume ($\approx 8.3$).  The remaining events (ToWall smaller than $200$~cm) are scaled by the ratio of the surface areas of the detectors ($\approx 3.6$). This difference in scaling is applied to take into account the fact that the event reconstruction performance is worse for interactions near the wall and the volume/surface ratio is different between Hyper-Kamiokande and Super-Kamiokande. The difference of the number of events between the ToWall dependent scaling and the full volume ratio scaling is on the order of $10\%$.

\begin{figure*}[th!]
\centering
        \includegraphics[width=0.48\textwidth]{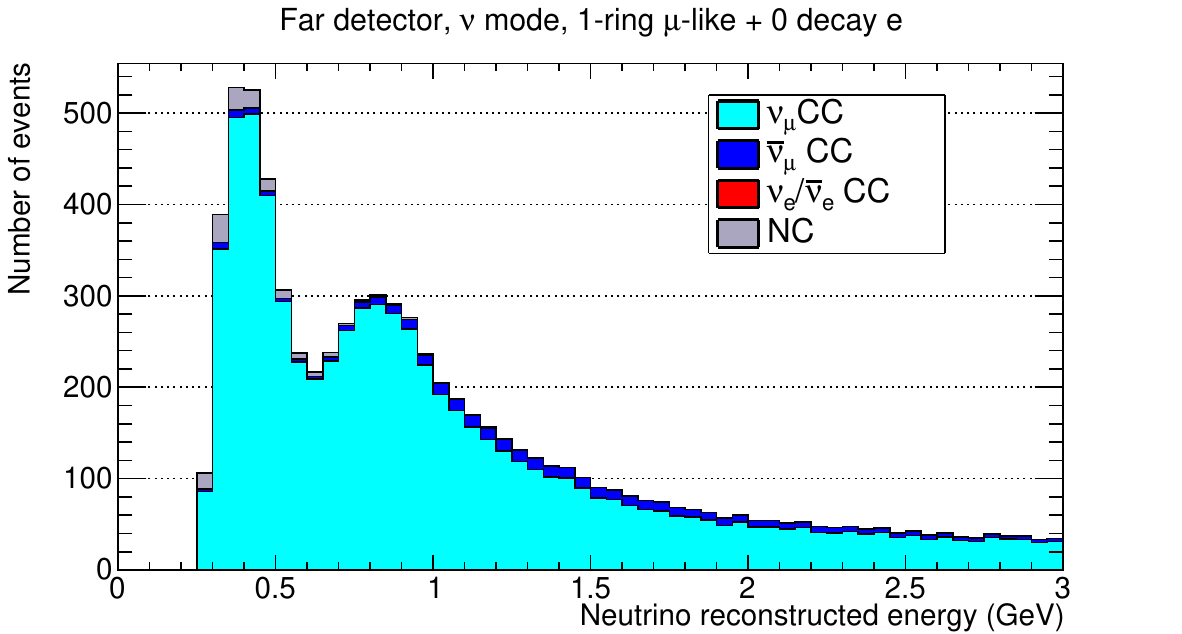}
        \includegraphics[width=0.48\textwidth]{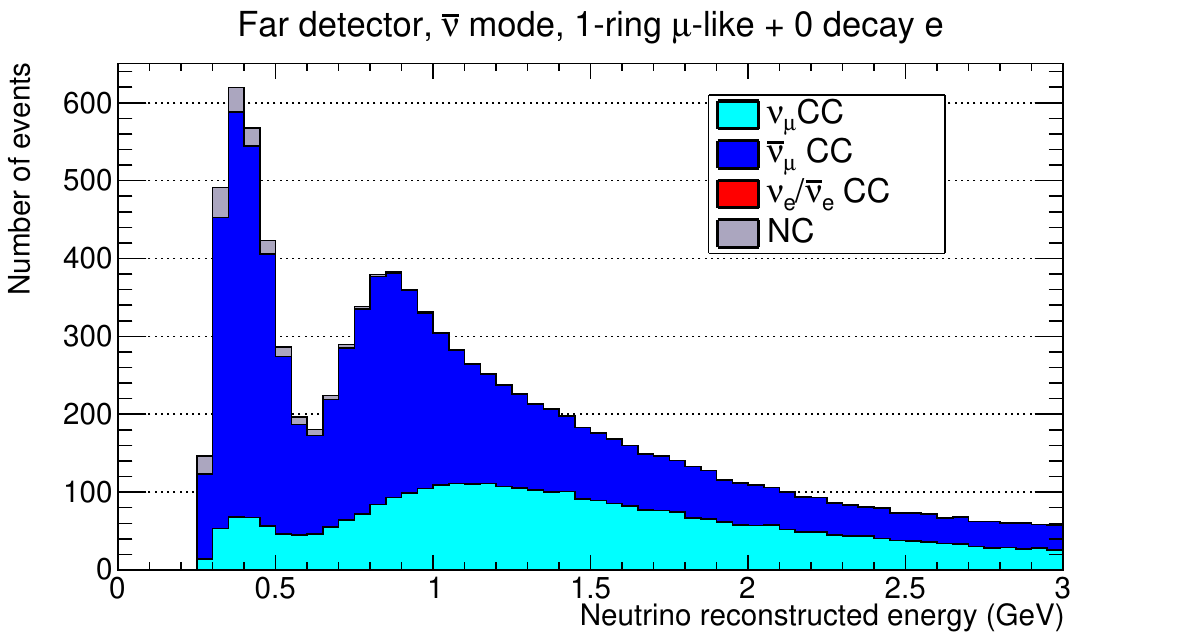}
        \includegraphics[width=0.48\textwidth]{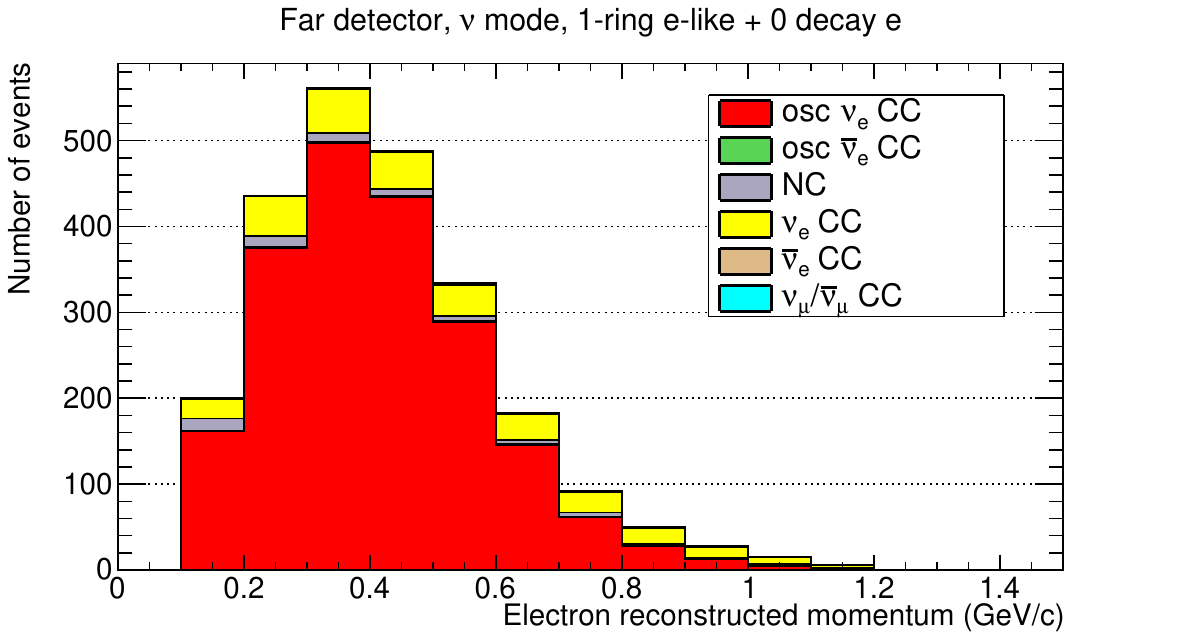}
        \includegraphics[width=0.48\textwidth]{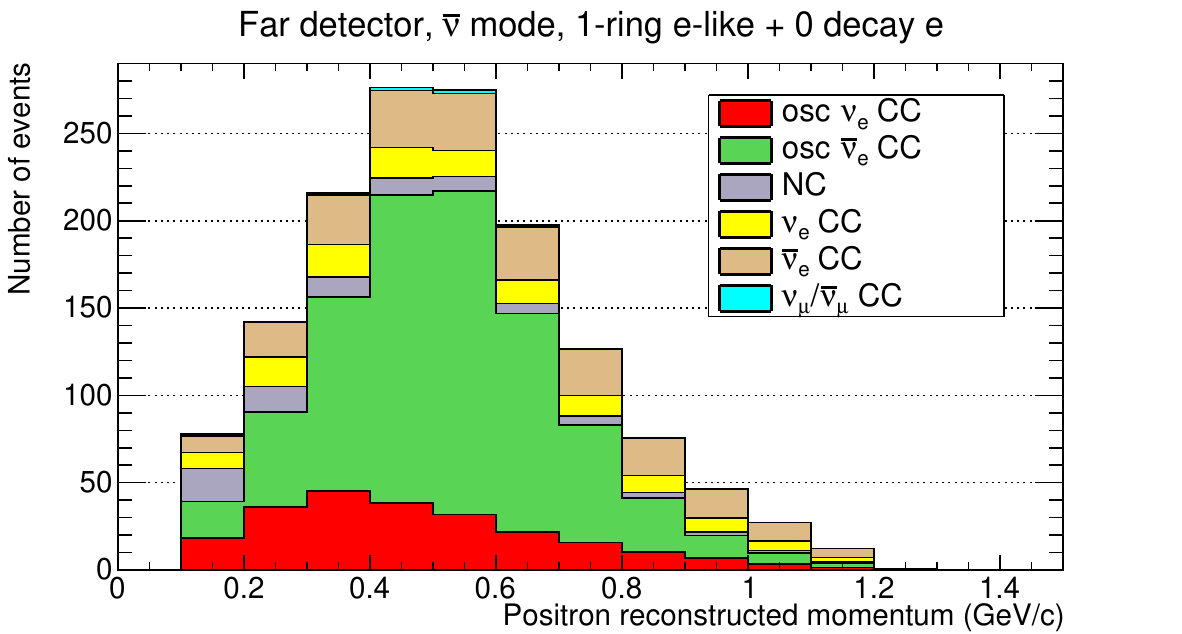}
        \includegraphics[width=0.48\textwidth]{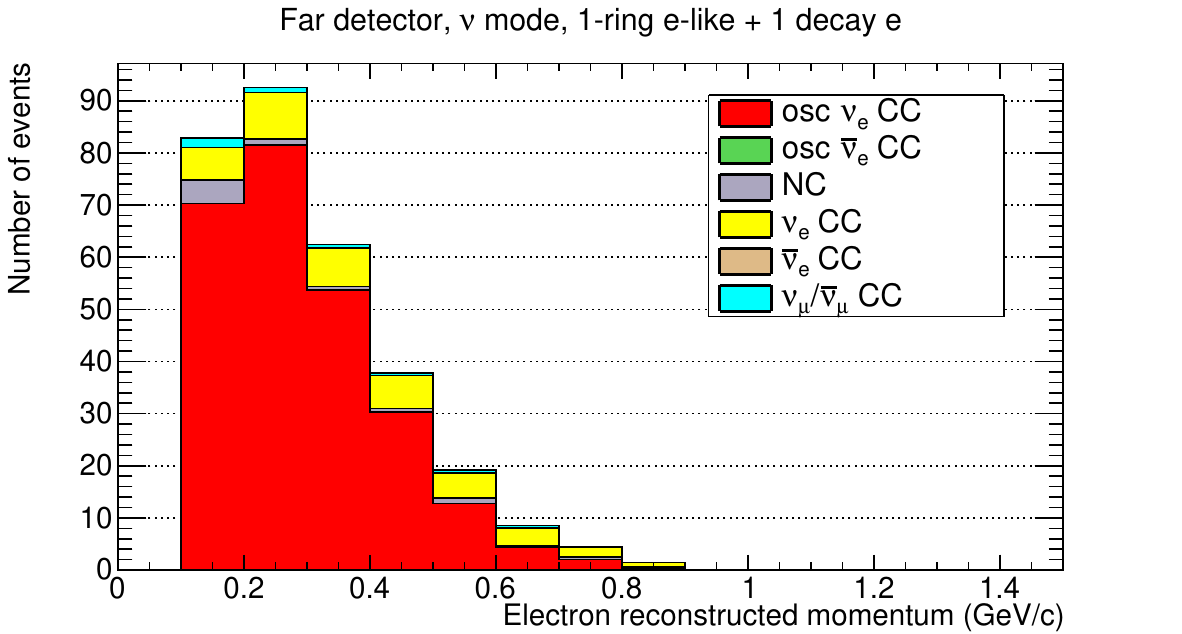}
    \caption{Reconstructed spectra of the selected samples predicted with $27\times10^{21}$ POT ($6.75\times10^{21}$ in FHC and $20.25\times10^{21}$ in RHC), corresponding to 10 years of accumulated statistics. The electron (anti)neutrino samples are separated between appearance neutrinos from oscillation (`osc') and the intrinsic electron neutrino component of the beam.}
    \label{fig:erec}
\end{figure*}

Unless specified otherwise, we will assume the value of oscillation parameters reported in Tab.~\ref{Table:AsimovA2020}. When relevant, the dependence of the Hyper-Kamiokande sensitivity to the specific value of these oscillation parameters will be studied.
We report in Tab.~\ref{Table:events} the number of expected events and their kinematic distributions are shown in Fig.~\ref{fig:erec}. The main kinematic variables considered in the analysis are the momentum and angle of the outgoing lepton: in the muon-like samples we use such observables to reconstruct the neutrino energy using the quasi-elastic assumption~\cite{T2K:2021xwb}, enabling more direct sensitivity to the shape of the oscillated energy spectrum. 

\begin{table*}[ht]
    \centering
    \begin{tabular}{c|c|c|c|c|c|c}
        $\sin^2\theta_{12}$ & $\Delta m^2_{21}$ & $\sin^2\theta_{23}$ & $\Delta m^2_{32}$ & $\sin^2\theta_{13}$ & $\delta_{CP}$  & mass ordering\\
        \hline
        0.307 & $7.53\times10^{-5}$eV$^2$ & 0.528 &  $2.509\times10^{-3}$eV$^2$ & 0.0218 & $-1.601$ rad & normal
    \end{tabular}
    \caption{Values of the oscillation parameters assumed for the sensitivity studies, unless specified otherwise.}
    \label{Table:AsimovA2020}
\end{table*}

\begin{table*}[ht]
    \centering

\begin{tabular}{c|c|c|c|c|c|c|c}
        ~ &	beam $\nu_\mu$ &	beam $\nu_e$ &	beam $\bar\nu_\mu$ &	beam $\bar\nu_e$ &	$\nu_\mu \to \nu_e$ &	$\bar\nu_\mu \to \bar\nu_e$  &	Total \\
\hline
$\nu$-mode, 1-ring $e$-like + 0 decay $e$ &	143.9 &	294.3 &	5.3 &	12.0 &	2007.5 &	11.7 &	2474.7 \\
$\bar\nu$-mode, 1-ring $e$-like + 0 decay $e$ &	59.1 &	130.1 &	96.3 &	234.8 &	229.2 &	793.2 &	1542.7 \\
$\nu$-mode, 1-ring $e$-like + 1 decay $e$ &	14.0 &	40.2 &	0.6 &	0.3 &	255.3 &	0.2 &	310.6 \\
\hline
$\nu$-mode, 1-ring $\mu$-like&	8355.4 &	8.4 &	478.0 &	0.7 &	2.6 &	0.01 &	8845.1 \\
$\bar\nu$-mode, 1-ring $\mu$-like &	4255.9 &	6.0 &	7759.9 &	4.7 &	0.2 &	0.4 &	12027.2 \\

    \end{tabular}
    \caption{Expected number of events at Hyper-Kamiokande with $27 \times 10^{21}$ POT ($6.75 \times 10^{21}$ POT in FHC and $20.25 \times 10^{21}$ POT in RHC), corresponding to 10 years of accumulated statistics. The first four columns correspond to non-oscillated events which include electron (anti)neutrinos from the intrinsic beam contamination.}
    \label{Table:events}
\end{table*}

\section{Analysis}
\label{sec:analysis}
\subsection{Overview}
The analysis developed for the T2K experiment~\cite{T2K:2023smv} is adapted to take into account the different configuration of beam and detectors in Hyper-Kamiokande. The general principles of the analysis are described below.

The uncertainty in the simulated neutrino flux and interaction cross-section models is parameterised. Then, the parameters are fit to near-detector event samples to both tune the model prediction and reduce the uncertainty on the predicted event spectra at the far detector.  This tuned model is then fit to the simulated far detector event samples to extract the sensitivity to the neutrino oscillation parameters, while also including uncertainties associated with the far detector reconstruction as nuisance parameters.

\subsection{Systematic uncertainties and near detector inputs}
\label{sec:systerrs}

We use the neutrino flux, neutrino-nucleus interaction cross-section and detector response models developed by the T2K collaboration and, therefore, adopt the same parameterisation of the model uncertainties.  A detailed description of these can be found in Ref.~\cite{T2K:2023smv}. 

The systematic uncertainties related to the modeling of the far detector are implemented as a weighting of the number of selected events according to the sample, neutrino interaction type and reconstructed neutrino energy. T2K has constrained these uncertainties using a fit to the Super-Kamiokande atmospheric neutrino samples and other dedicated background control samples.

The flux systematic uncertainties are implemented in terms of binned nuisance parameters which weight the number of expected neutrinos at Hyper-Kamiokande as a function of neutrino type (electron or muon flavour for neutrinos and antineutrinos), beam mode (FHC and RHC) and neutrino energy. Uncertainties are evaluated using NA61/SHINE hadroproduction measurements, beam line modeling and alignment uncertainties, measurements of the horn current, proton beam monitoring data, and measurements from the on-axis neutrino beam monitor INGRID. We assume that Hyper-Kamiokande will achieve a level of beam systematic uncertainties similar to that T2K has demonstrated.

We consider four main types of neutrino-nucleus interactions: charged-current quasi-elastic, interactions with pairs of correlated nucleons (also called 2p2h), single pion production and other interactions (including multi-pion production and deep inelastic scattering). The kinematics, type and number of particles observed in the detector can be further modified by so-called ``final state interactions'' (FSI) of pions and nucleons as they exit the nucleus. Coulomb corrections to the momenta of charged particles leaving the nucleus are also implemented. Uncertainties on the fundamental physics parameters inside the interaction model are included when technically possible, or otherwise as effective weights in bins of the fundamental kinematic variables (neutrino energy, transferred 4-momentum to the nucleus, among others) or as overall normalization uncertainties for specific processes. These uncertainties are set based on theoretical arguments and neutrino cross-section measurements from the T2K near detectors and other dedicated experiments. 

The neutrino flux and cross-section uncertainties will be constrained in Hyper-Kamiokande by a set of near detectors, including the upgraded T2K near detector ND280 and a new Intermediate Water Cherenkov Detector. In this analysis, we consider the constraints obtained by the T2K experiment from ND280 data in Ref.~\cite{T2K:2023smv}, and we assume further improvements on the basis of the expected increase of statistics at Hyper-Kamiokande and of the improved capabilities of upgraded ND280~\cite{Dolan:2021hbw} and IWCD.  Table~\ref{tab:Onesigma_VALOR_t2K} shows the resulting systematic uncertainties on the number of expected events at Hyper-Kamiokande.

\begin{table*}[ht]
    \centering
    \begin{tabular}{c|c|c|c|c|c|c}
    \hline
        T2K systematics & FHC 1Re & FHC 1R$\mu$ & RHC 1Re & RHC 1R$\mu$ & FHC 1Re1De & FHC/RHC 1Re\\
         \hline 
        Flux-xsec & $3.6\%$ & $2.1\%$ & $4.3\%$ & $3.4\%$ & $4.9\%$ & $4.4\%$ \\ 
        Detector & $3.1\%$ & $2.1\%$ & $3.9\%$ & $1.9\%$ & $13.2\%$ & $1.1\%$ \\
        All & $4.7\%$ & $3.0\%$ & $5.9\%$ & $4.0\%$ & $14.1\%$ & $4.6\%$\\
\hline\hline
     Improved systematics & FHC 1Re & FHC 1R$\mu$ & RHC 1Re & RHC 1R$\mu$ & FHC 1Re1De & FHC/RHC 1Re\\
      \hline 
        Flux-xsec & $1.8\%$ & $0.9\%$ & $1.6\%$ & $0.9\%$ & $1.8\%$ & $1.9\%$\\
        Detector & $1.1\%$ & $0.8\%$ & $1.5\%$ & $0.7\%$ & $4.9\%$ & $0.4\%$ \\
        All & $2.1\%$ & $1.2\%$ & $2.2\%$ & $1.1\%$ & $5.2\%$ & $2.0\%$ \\
\hline
    \end{tabular}
\caption{$1\sigma$ uncertainty on the number of events expected in each sample for each source of uncertainty for either the same systematic errors as the T2K analysis~\cite{T2K:2023smv} after the near detector fit (T2K syst.) or an improved error model considering 10 years of data accumulated at the near and far detectors (Impr. syst.). The uncertainty on the ratio of 1Re event sample in FHC and RHC is reported in the last column.}
    \label{tab:Onesigma_VALOR_t2K}
\end{table*}

We build the improved systematics prediction by modifying the errors associated with each systematic parameter in the T2K model without modifying the correlations between the parameters.
For all parameters constrained by ND280, the uncertainty is scaled by $\sqrt{1/N}$, considering the relative increase in beam exposure between Hyper-Kamiokande and T2K ($3.6\times 10^{21}$~POT), corresponding to about a factor of 7.5 in POT for the 10 years of Hyper-Kamiokande. 
This makes the implicit assumption that the final uncertainty from the near detector constraint is limited by the statistics at the near detector and, as a consequence, the near detector systematic uncertainties must be smaller than the statistical error. This assumption is verified in T2K, it is further ensured by the upgraded ND280 and is fundamentally  motivated by the fact that systematic uncertainties can be constrained from control samples which increase in statistics at the same pace of the signal samples.
We also assume that the near to far detector data-taking ratio will be similar between T2K and Hyper-Kamiokande. Finally, the far detector systematic uncertainties from Ref.~\cite{T2K:2023smv} are also scaled by the same factor $\sqrt{7.5}$ considering 10 years of Hyper-Kamiokande.
This is a somewhat arbitrary assumption: the far detector systematic uncertainties will be constrained using calibration sources and various control samples.  We assume that the achievable precision will roughly scale with the collected statistics, which increases with the volume change between Hyper-Kamiokande and Super-Kamiokande.

In addition to the reduction due to increased statistics, further reductions to individual parameter uncertainties were applied based on the expected performance of the upgraded ND280 and IWCD detectors. In general, such detectors will feature improved angular acceptance and much increased target mass. The improved reconstruction of the hadronic final state in the upgraded ND280, notably enabling lower threshold for proton reconstruction, is expected to strongly reduce the quasi-elastic uncertainties~\cite{Dolan:2021hbw}. In addition, lower pion threshold will further improve the precision of pion-production measurements. In case of neutron production, dominant in antineutrino interactions, the upgraded ND280 will allow for the first time the measurement of neutron kinematics but with somewhat lower efficiency and less precision than for final states with protons~\cite{Munteanu:2019llq}. IWCD will provide extremely large statistics samples for both charged and neutral current interactions.  The off-axis spanning capability of IWCD provides a direct link between neutrino energy and reconstructed particle kinematics, allowing precise measurements of energy mis-reconstruction and as a result improved constraints on charged current systematic uncertainties, particularly those associated to multi-nucleon and resonant interactions.  Water Cherenkov detectors also provide high purity and efficiency samples of neutral current interactions through the reconstruction of neutral pions.  On the basis of these considerations, the following reductions of systematic uncertainties have been applied. For charged-current interactions, we have reduced neutrino non-quasi-elastic uncertainties by a factor of three, quasi-elastic uncertainties by a factor of 2.5 and all antineutrino uncertainties by a factor of 2. Uncertainties on neutral current interactions were reduced to $\sim10\%$. The energy dependent uncertainties on 2p2h are not modified: while the off-axis spanning of IWCD and the improved energy reconstruction capabilities of the upgraded ND280 are expected to improve energy-dependent uncertainties, we defer quantitative evaluation to further studies. 
The errors on the $\sigma(\nu_e)/\sigma(\nu_\mu)$ and $\sigma(\bar{\nu}_e)/\sigma(\bar\nu_\mu)$ cross-section ratios are fixed so that the error on the ratio is $2.7\%$ (unless specified otherwise) and the correlation between the two parameters is $-0.33$, as in the T2K model~\cite{Abe:2017vif}.
Theoretical uncertainties on the $\nu_e/\bar{\nu}_e$ interaction cross-section, with respect to the $\nu_{\mu}/\bar{\nu}_{\mu}$ cross-section, arise from mis-modeling of nuclear effects and radiative corrections, which depend on the lepton mass difference between electrons and muons in charged-current interactions. 
The systematic parameters governing the magnitude of these two effects are mostly correlated between $\nu$ and $\bar{\nu}$.
The impact of $\nu_e/\bar{\nu}_e$ uncertainty in the Hyper-Kamiokande flux and kinematic region is studied in Ref.~\cite{Dieminger:2023oin} for the nuclear physics uncertainties and Ref.~\cite{Tomalak:2021hec} for the radiative corrections. 
Both papers study the specific Hyper-Kamiokande kinematic region showing, respectively, that nuclear uncertainties are expected to be below $2\%$ and a kinematic-dependent prediction of radiative corrections gives a residual uncertainty below 0.5\%.
Such theoretical inputs will be further corroborated by direct measurements at IWCD and upgraded ND280. The $2.7\%$ uncertainty on $\sigma(\nu_e)/\sigma(\bar\nu_e)$ is thus a reasonable target in the Hyper-Kamiokande era.

The model of uncertainties developed by the T2K collaboration is very detailed and its robustness has been proven in multiple iterations of T2K data analysis and in dedicated simulated test datasets built using alternative neutrino interaction models. While such a model will certainly be further refined in anticipation of the increased statistical power of both the ongoing T2K and upcoming Hyper-Kamiokande experiments, it is a reasonable,
data-driven approach to test the sensitivity of Hyper-Kamiokande.
The robustness of the results for very large statistics 
was tested in various ways. We enforced each single systematic uncertainty to be constrained not better than 1\% and obtained stable results, ensuring that they are not dominated by any over-constrained uncertainty. We also tested that in the case of maximal (10 years) far detector statistics, the primary constraint on the model is still coming from the near detectors.
Note that unless specified otherwise, the improved systematics prediction is built considering the expected statistical power after $10$ years of Hyper-Kamiokande operation.

As a comparison, we also report sensitivities for Hyper-Kamiokande that use the T2K uncertainties as in Ref.~\cite{T2K:2023smv}, as well as sensitivities assuming statistical errors only.  These results highlight the impact that the systematics error model has on the physics reach of Hyper-Kamiokande. 
 
\subsection{Fit of the oscillation parameters}

The neutrino oscillation parameter sensitivities reported here assume a full three-flavor PMNS parameterization of neutrino mixing.  The oscillation parameters \(\delta_{CP}\), \(\theta_{23}\), and \(\Delta m^2_{32}\) are fit without external constraints, while \(\theta_{13}\) is fit both without and with a Gaussian external constraint of \(\sin^22\theta_{13} = 0.0853 \pm 0.0027\) coming from measurements using reactor antineutrinos \cite{ParticleDataGroup:2024cfk}.  The values of $\sin^2\theta_{12}=0.307$ and $\Delta m^2_{21}=7.53 \times 10^{-5}$~eV$^2$ are held fixed in the fit following the measurements from solar and reactor experiments~\cite{ParticleDataGroup:2024cfk}. 

The analysis is based on a binned maximum likelihood method where the likelihood is defined as:
\begin{equation}
\begin{split}
    &\mathcal{L}(\{N^\text{obs}_s, \mathbf{x}^\text{obs}_s\}_{\forall s}, \mathbf{o}, \mathbf{f})= \\
    &\prod_{s\in\text{samples}}[\mathcal{L}_s(N^\text{obs}_s, \mathbf{x}^\text{obs}_s, \mathbf{o}, \mathbf{f})]\times  \mathcal{L}_\text{syst}(\mathbf{f}),
\end{split}
\end{equation}
where $s$ runs through the samples considered. $N^\text{obs}_s$ is the number of candidate events observed for sample $s$ and $\mathbf{x}^\text{obs}_s$ represent the measurement variables: the electron-like samples are binned into reconstructed charged lepton momentum and scattering angle, while the muon-like samples are binned into reconstructed neutrino energy and charged lepton scattering angle. The symbol $\mathbf{o}$ represents the set of all oscillation parameters we measure, and $\mathbf{f}$ is the set of systematic nuisance parameters.
$\mathcal{L}_\text{syst}(\mathbf{f})$, the term describing the systematic effects: 
\begin{equation}
\mathcal{L}_\text{syst}(\mathbf{f}) = \exp{\left(-\frac{(\mathbf{f}-\mathbf{f_0})^{T}V^{-1}(\mathbf{f}-\mathbf{f_0})}{2} \right)},
\end{equation}
where $\mathbf{f_0}$ is the set of prior preferred values of the systematic parameters  and $V$ is the covariance matrix that describes the input uncertainty on the systematic parameters and their correlations. 
This frequentist analysis is performed with the same fitting framework of Ref.~\cite{T2K:2023smv} and all nuisance (systematic and oscillation) parameters are profiled.

%\newpage
%\newpage
\section{Results}
\label{sec:results}
\begin{figure}[ht!]
        \includegraphics[width=0.5\textwidth]{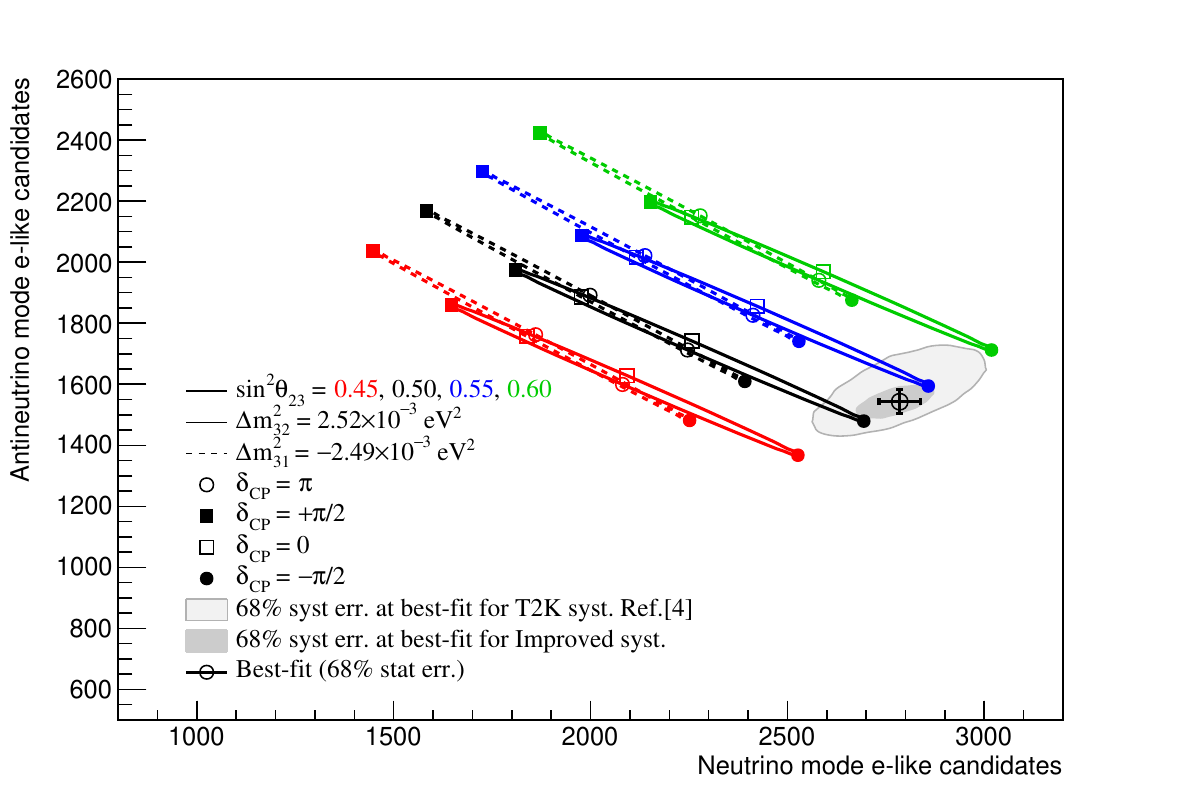}\caption{Fit results with statistical and systematic uncertainties projected on the number of electron (anti)neutrino candidates after 10 years of data taking. The expected number of events for different values of the oscillation parameters is shown for comparison.}
    \label{fig:biprob}
\end{figure}

\begin{figure*}[ht!]
\centering
        \includegraphics[width=0.48\textwidth]{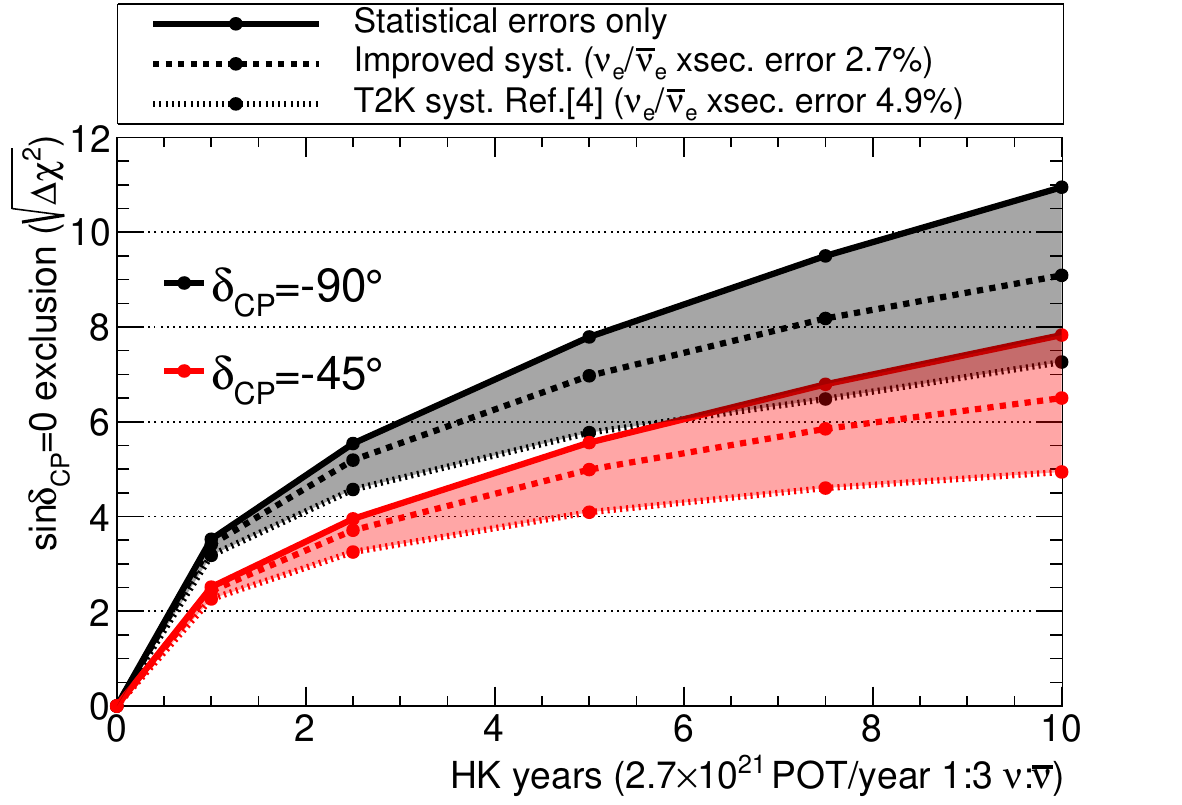}
        \includegraphics[width=0.48\textwidth]{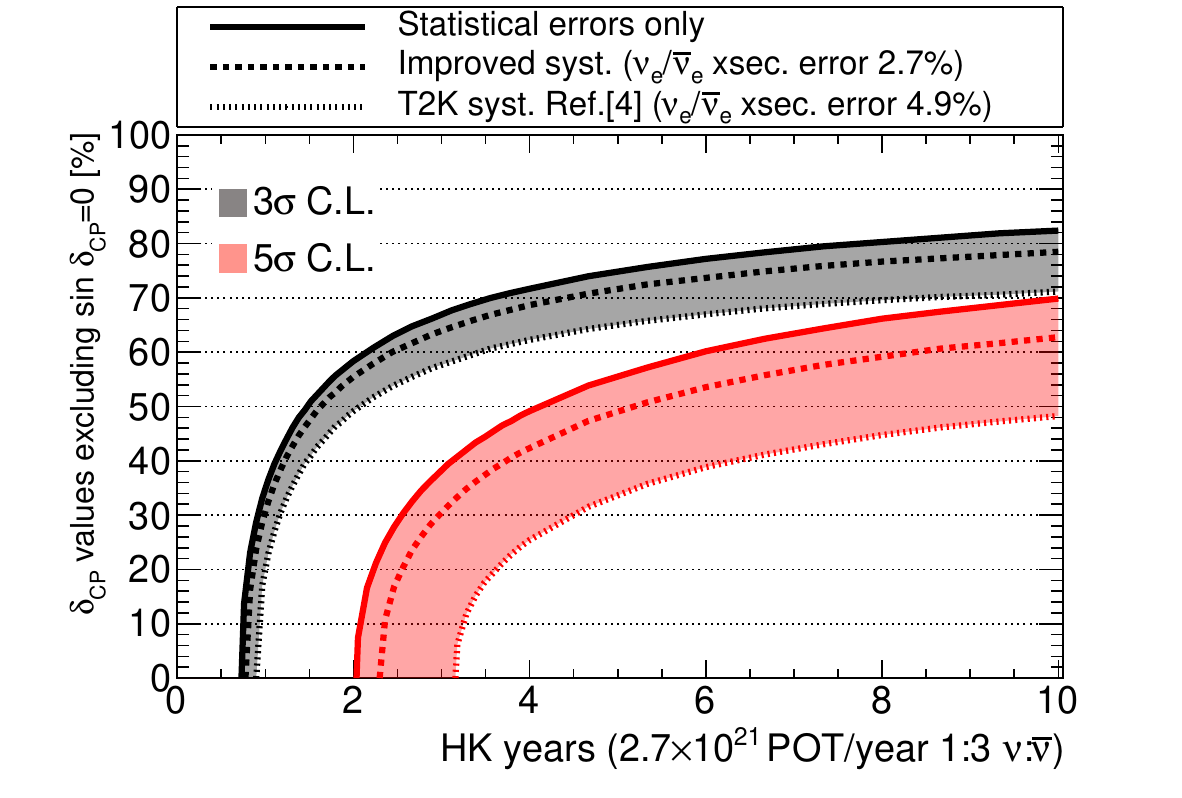}      
    \caption{Sensitivity to CPV as a function of data-taking time:
    $\sin\delta_{CP}=0$ exclusion for $\delta_{CP}=-90^\circ$ or $-45^\circ$ (left) and percentage of $\delta_{CP}$ values for which $\sin\delta_{CP}=0$ can be excluded at 3$\sigma$ and at 5$\sigma$ (right). The shaded regions in these and following figures, show the span of possible sensitivities when varying the assumed systematic errors.}
    \label{fig:cpv}
\end{figure*}

The results of the fit to the oscillation parameters, on a sample corresponding to 10 years of Hyper-Kamiokande data, is projected in Fig.~\ref{fig:biprob} on the number of electron (anti)neutrino candidates, the most relevant samples for the CPV search. The best fit and its statistical and systematic uncertainties are compared to the expected number of events for different values of the oscillation parameters.
The Hyper-Kamiokande sensitivity for CPV discovery as a function of time is shown in Fig.~\ref{fig:cpv}.
In the case of maximal CPV, $\delta_{CP}=-\pi/2$, Hyper-Kamiokande reaches a definitive $5\sigma$ discovery in less than three years. Even with a very conservative assumption on the systematic uncertainties (same values as in T2K in Ref.~\cite{T2K:2023smv}), Hyper-Kamiokande will reach $5\sigma$ sensitivity in less than six years. Given the extremely fast discovery scenario for such a case, the possibility of a still partially unknown mass ordering (MO) should be considered.
If we consider the degenerate case of $\delta_{CP}=-\pi/2$ and inverted ordering and the Hyper-Kamiokande sensitivity to MO using atmospheric and beam data~\cite{Hyper-Kamiokande:2018ofw}, the CPV discovery would be delayed to six years also in the case of improved systematic uncertainties. Such estimation is obtained by considering the MO determination with atmospheric Hyper-Kamiokande data as an external constrain to the CPV search with Hyper-Kamiokande beam data. This is a conservative estimate, since it does not consider external measurements of MO from other experiments, nor the boost in CPV sensitivity obtained by a full joint fit of beam and atmospheric data in Hyper-Kamiokande, as performed for instance by the T2K and Super-Kamiokande collaborations in Ref.~\cite{T2K:2024wfn}. Such joint Hyper-Kamiokande beam-atmospheric analysis is being performed for Hyper-Kamiokande and its sensitivity will be reported in a future paper.

If CP is not maximally violated, assuming known MO, after about six years, Hyper-Kamiokande will be able to discover CPV at 3$\sigma$ (5$\sigma$) for 75\% (55\%) of possible actual values of $\delta_{CP}$. For instance, a CPV discovery could be achieved in under six years in case of $\delta_{CP}=-\pi/4$. The fraction of possible actual values of $\delta_{CP}$ for which Hyper-Kamiokande can discover CPV at 3$\sigma$ (5$\sigma$) with 10-years exposure is about 80\% (60\%). These fractions of $\delta_{CP}$ values increase (decrease) by few percent by changing the value of $\sin^2\theta_{23}$ to 0.4 (0.6) and they are independent on the actual MO. The systematic error which most degrades the sensitivity of Hyper-Kamiokande to CPV is the uncertainty on the \(\sigma(\nu_e)/\sigma(\bar{\nu}_e)\) cross-section ratio.

\begin{figure*}[ht!]
    \centering
        \includegraphics[width=0.48\textwidth]{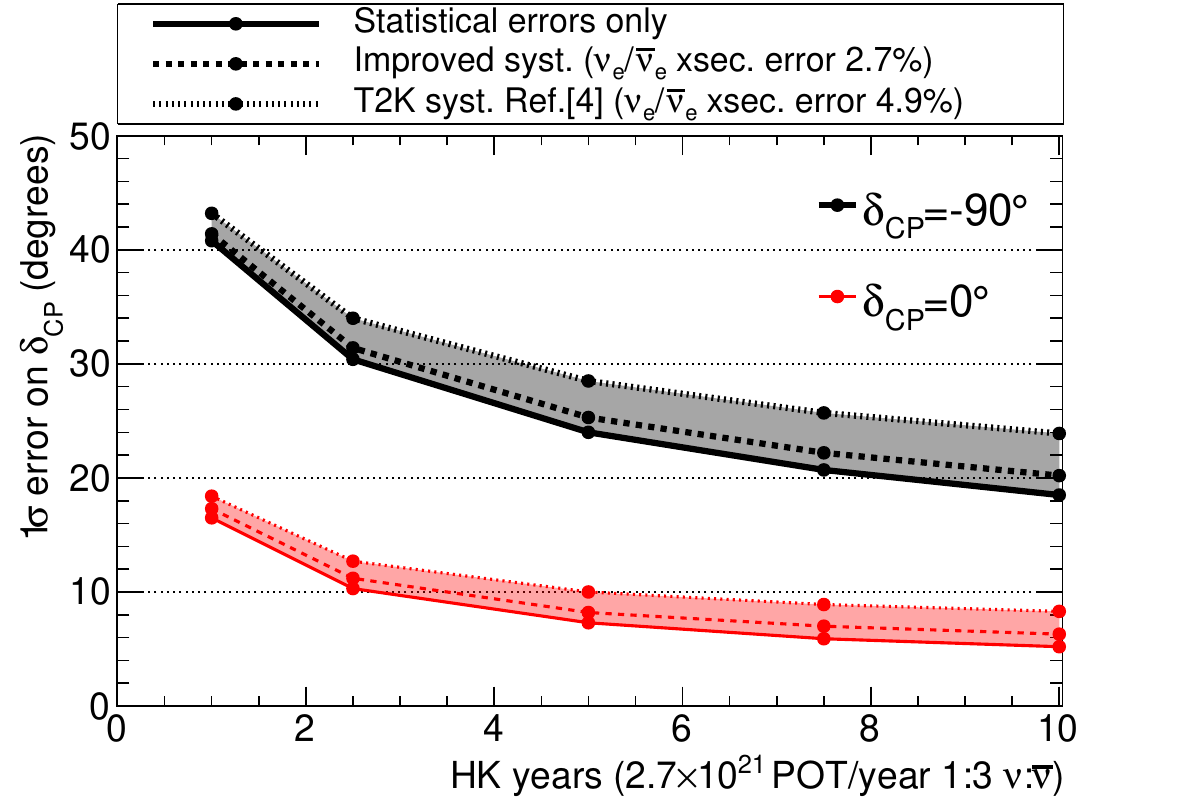}
        \includegraphics[width=0.48\textwidth]{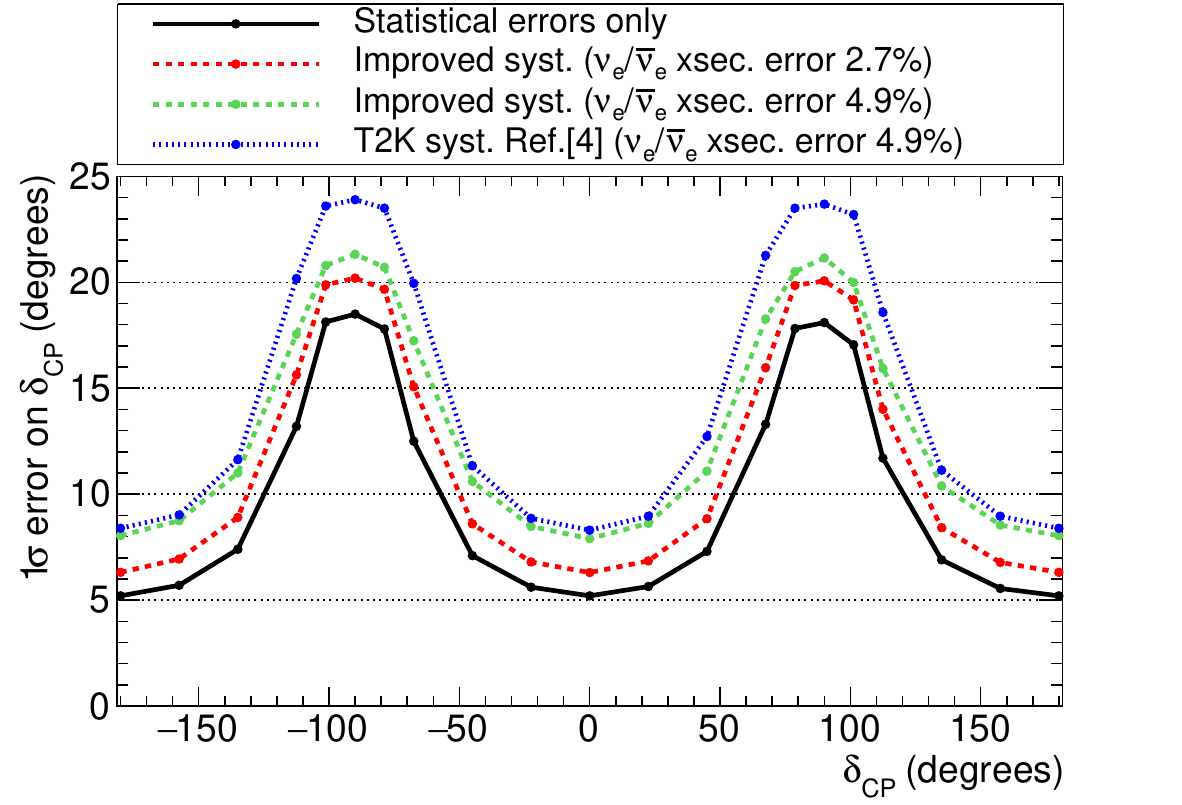}
 \caption{$1\sigma$ error on $\delta_{CP}$ as a function of data-taking time assuming $\delta_{CP}=-\pi/2$ or $0$ (left) and as a function of the value of $\delta_{CP}$ after 10 years of data taking (right). Results with different uncertainties on $\sigma(\nu_e)/\sigma(\bar\nu_e)$ are shown. }
    \label{fig::res_dcp}
\end{figure*}

The expected resolution of the $\delta_{CP}$ measurement is shown in Fig.~\ref{fig::res_dcp}. The achievable resolution depends on the actual value of $\delta_{CP}$, where better resolution can be achieved for values of \(\delta_{CP}\) close to those where CP is conserved. The most relevant systematic uncertainties are also different depending on the actual value of $\delta_{CP}$. In Fig.~\ref{fig::res_dcp}, two scenarios for the improved systematic error model with different constraints on $\sigma(\nu_e)/\sigma(\bar\nu_e)$ are tested. To evaluate the resolution achievable for the measurement of $\delta_{CP}$, we have to consider the derivative with respect to $\delta_{CP}$ of the oscillation formula in Eq.~\ref{eq:Pap}: 
\begin{multline}
 \partial P(\nu_\mu \rightarrow \nu_e)/\partial \delta_{CP} = \\
 - 8c_{13}^2c_{12}c_{23}s_{12}s_{13}s_{23}\cos\delta_{CP}\cdot \sin\Delta_{32}\cdot \sin\Delta_{31}\cdot \sin\Delta_{21}\\
  +8c_{13}^2s_{12}s_{13}s_{23}c_{12}c_{23}\sin\Delta_{21}\sin\delta_{CP}\\(s^2_{12}\sin\Delta_{21}-\cos\Delta_{32}\sin\Delta_{31}).
 \label{eq:dPap}
 \end{multline}

For the case of CP-conservation ($\delta_{CP}=0,\pi$), the CP-odd term ($\sin\delta_{CP}$) goes to 0 but its derivative ($\cos\delta_{CP}$) is maximal, thus the precision on the $\delta_{CP}$ measurement is dominated by this CP-odd term. Therefore, the precision measurement of $\delta_{CP}$ around the CP-conserving values is mostly a rate measurement of the difference between $\nu_e$ and $\bar{\nu}_e$, thus the $\sigma(\nu_e)/\sigma(\bar\nu_e)$ ratio has a significant impact on the resolution, as is the case for the CP-violation discovery sensitivity. For the case of maximal CP-violation, the situation is opposite: the CP-even term ($\cos\delta_{CP}$) goes to 0 while its derivative ($-\sin\delta_{CP}$) is maximal, thus dominating the resolution on the $\delta_{CP}$ measurement. In this case, the precision is not dominated by the rate asymmetry in the (anti)neutrino electron appearance channels, but by $\cos\delta_{CP}$-induced shape effects on their energy spectra, thus making a precise measurement more challenging in terms of statistics and requiring very good control of systematic effects related to the neutrino reconstructed energy, such as the far detector energy scale. As a consequence, in this case the $\sigma(\nu_e)/\sigma(\bar\nu_e)$ uncertainty does not dominate, and other systematic uncertainties have an important impact on the measurement.

\begin{figure*}[ht!]
    \centering
    \includegraphics[width=0.48\textwidth]{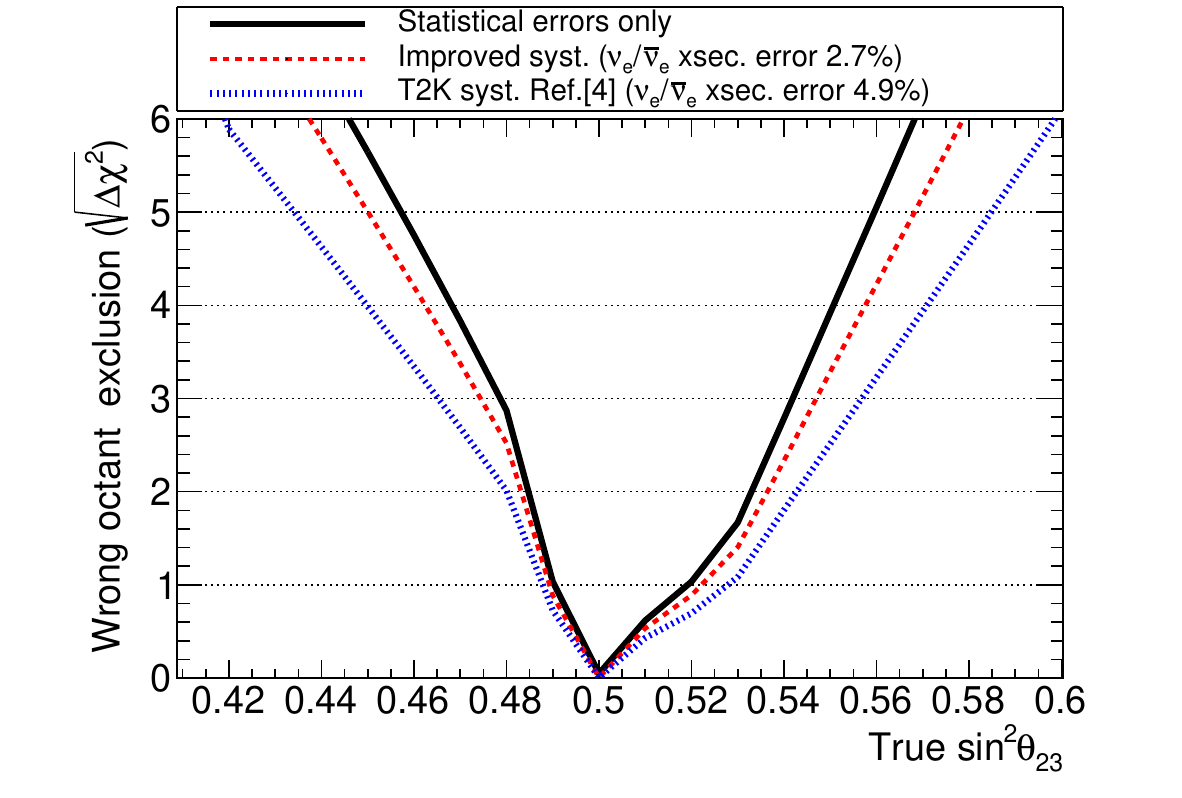}
    \includegraphics[width=0.48\textwidth]{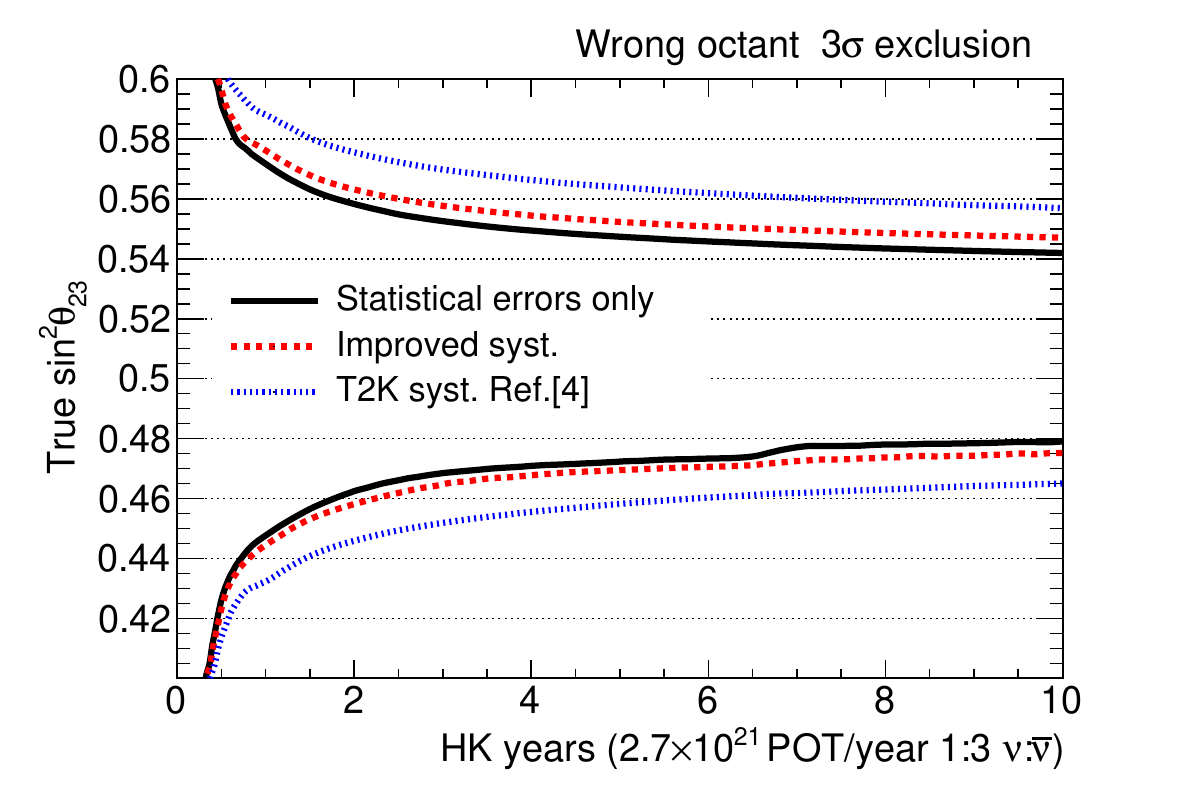}
    \caption{Sensitivity to the wrong \(\theta_{23}\) octant exclusion as a function of \(\theta_{23}\) value after 10 years of data taking (left). \(\theta_{23}\) region, as a function of data-taking time, for which 3\(\sigma\) exclusion of the wrong \(\theta_{23}\) octant can be reached (right).}
    \label{fig:wr_10years_allsyst}
\end{figure*}

Beyond the search for CPV, Hyper-Kamiokande will also feature unprecedented precision on the so-called atmospheric neutrino oscillation parameters.
The sensitivity to exclude the wrong $\theta_{23}$ octant is defined as 
\begin{equation}
\sqrt{\chi_{min}^2(\sin^2\theta_{23})_{WO}-\chi_{min}^2(\sin^2\theta_{23})_{RO}}
\end{equation}
where the labels $WO$ and $RO$ refer, respectively, to the wrong and right octant.
The Hyper-Kamiokande sensitivity to exclude the wrong $\theta_{23}$ octant is shown in Fig.~\ref{fig:wr_10years_allsyst} and summarized in Table~\ref{tab:octant}.

\begin{table}
\begin{tabular}{ c c c }
\hline \hline
 & C.L. $3 \sigma$ & C.L. $5 \sigma$\\
 \hline
Stat. only &  $[0,0.48] \cup [0.54,1]$ & $[0,0.46] \cup [0.56,1]$\\
Improved syst. & $[0,0.47] \cup [0.55,1]$ & $[0,0.45] \cup [0.57,1]$\\
T2K syst. in~\cite{T2K:2023smv} & $[0,0.46] \cup [0.56,1]$ & $[0,0.43] \cup [0.59,1]$\\
\hline \hline
\end{tabular}
\caption{Regions of values of $\sin^2\theta_{23}$ for which an exclusion of the wrong octant at 3$\sigma$ or 5$\sigma$ is possible after 10 years of data taking. }
\label{tab:octant}
\end{table}

\begin{figure*}[ht!]
    \centering
     \includegraphics[width=0.48\textwidth]{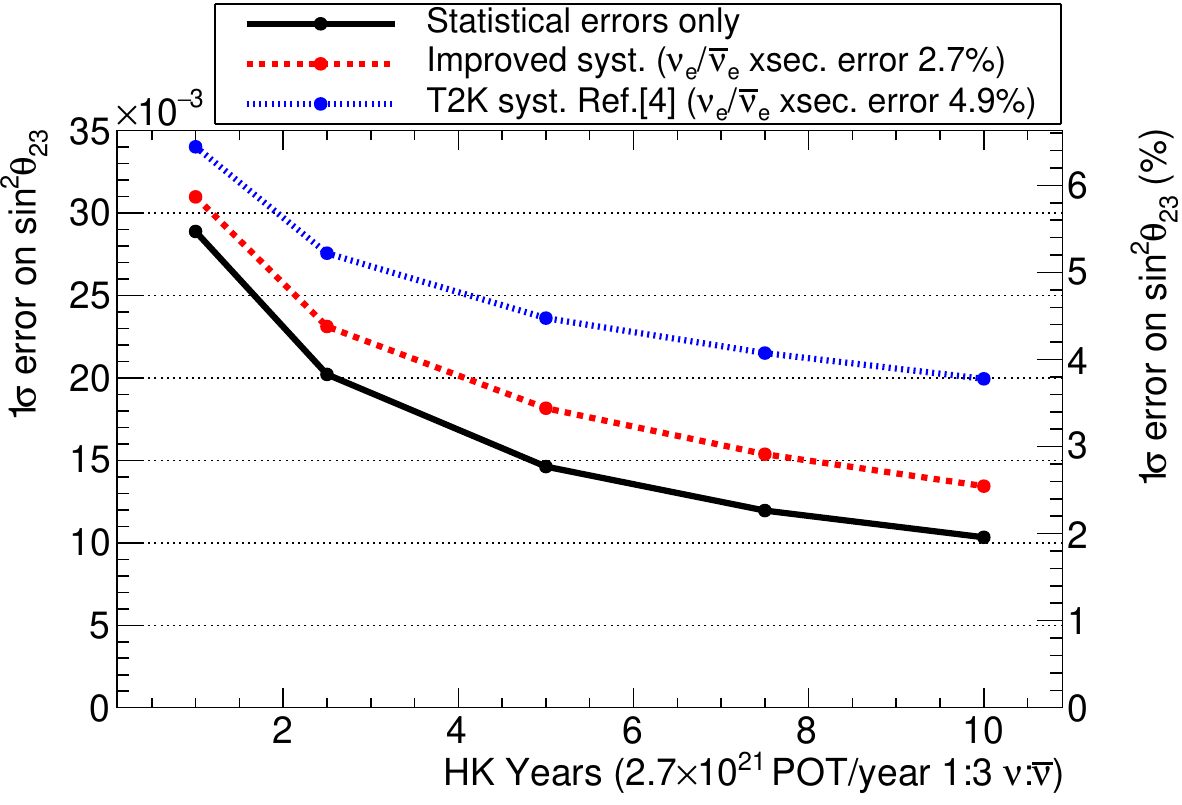}
    \includegraphics[width=0.48\textwidth]{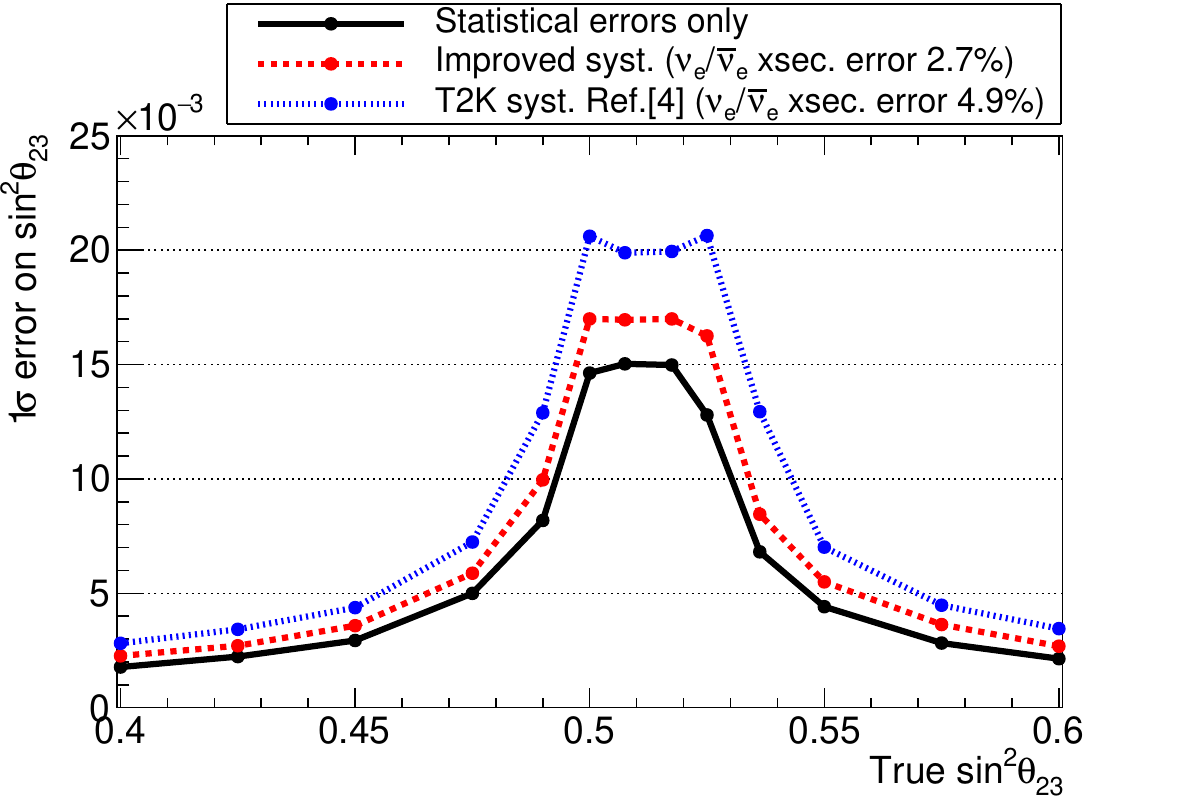}
    \caption{$1\sigma$ error on $\sin^2\theta_{23}$ as a function of data-taking time for $\sin^2\theta_{23}=0.528$  (left) and as a function of $\sin^2\theta_{23}$ value after 10 years of data taking (right).}
    \label{fig::res_t23}
\end{figure*}

Fig.~\ref{fig::res_t23} shows the resolution on the $\sin^2\theta_{23}$ mixing parameter achievable by Hyper-Kamiokande. Depending on the actual value of the parameter, an ultimate resolution between 2\% and 0.4\% can be reached.  The most challenging region is the case of maximal mixing, where the derivative of the oscillation probability is small and the resolution is directly affected by the octant degeneracy. 

\begin{figure}[ht!]
    \centering
    \includegraphics[width=0.48\textwidth]{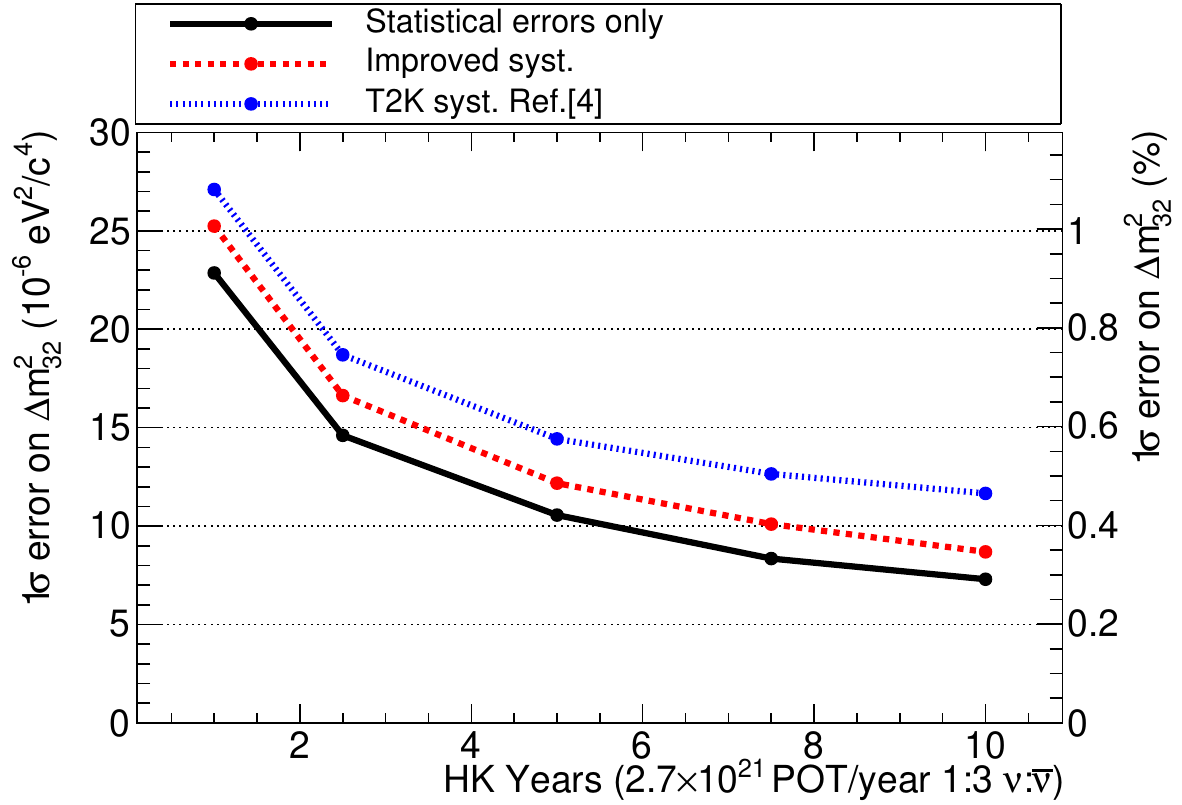}
    \caption{$1\sigma$ error on $\Delta m_{32}^2$ as a function of data-taking time.}
    \label{fig::res_dm32}
\end{figure}

The ultimate resolution achievable by Hyper-Kamiokande on $\Delta m_{32}^2$ is around 0.4\%, as shown in Fig.~\ref{fig::res_dm32}. This resolution does not depend sizeably on the actual value of the oscillation parameters. Reaching such precision will require an extremely robust model of systematic uncertainties, notably considering the detector energy scale calibration and the constraint on the nuclear removal energy. In turn, such extremely precise measurement of $\Delta m_{32}^2$ has important consequences in joint fits with reactor measurements for the determination of the MO~\cite{Nunokawa:2005nx,Parke:2024xre}.  

\begin{figure}[h!]
    \centering 
    \includegraphics[width=0.48\textwidth]{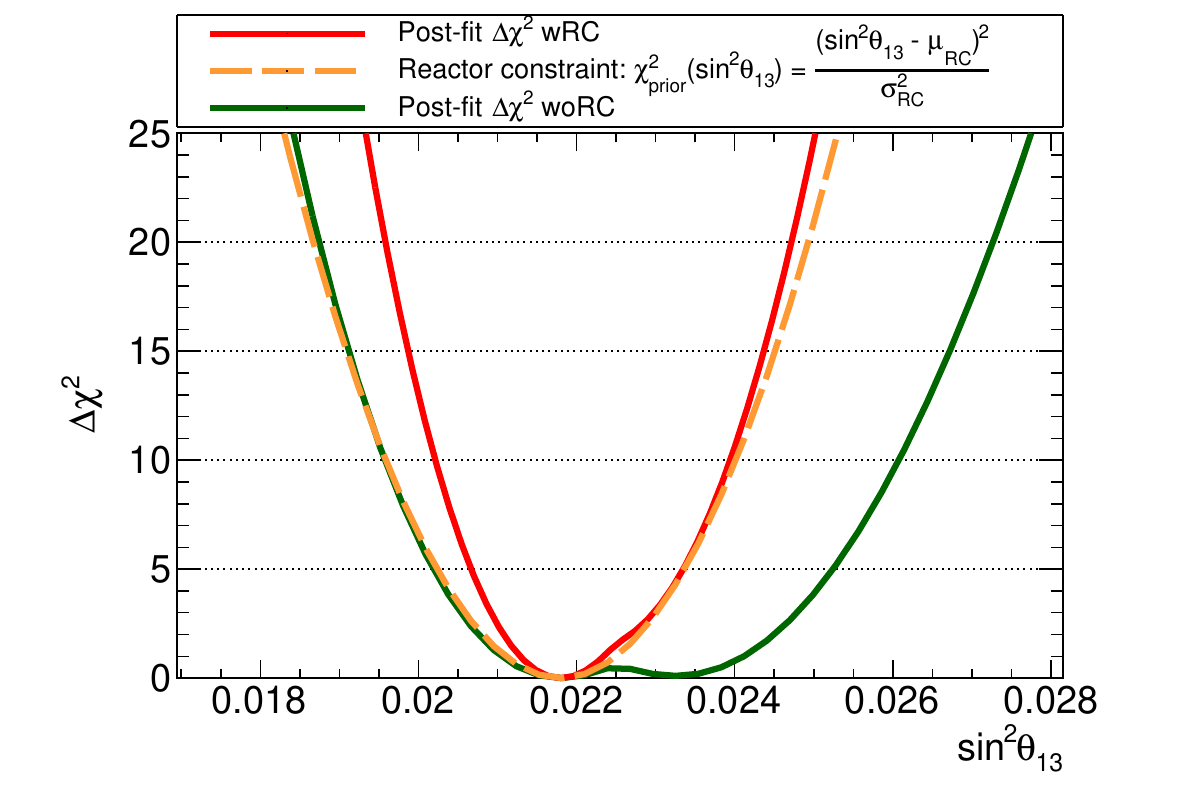}
    \caption{Measurement of \(\theta_{13}\): $\Delta \chi^2(\sin^2 \theta_{13})$ curves in the ``Improved syst.'' error model, after 10 years of data-taking, with (wRC) and without (woRC) the external constraint from reactor measurements.}
    \label{fig:s13_wrc_worc}
\end{figure}

\begin{figure}[ht]
    \centering 
    \includegraphics[width=0.5\textwidth]{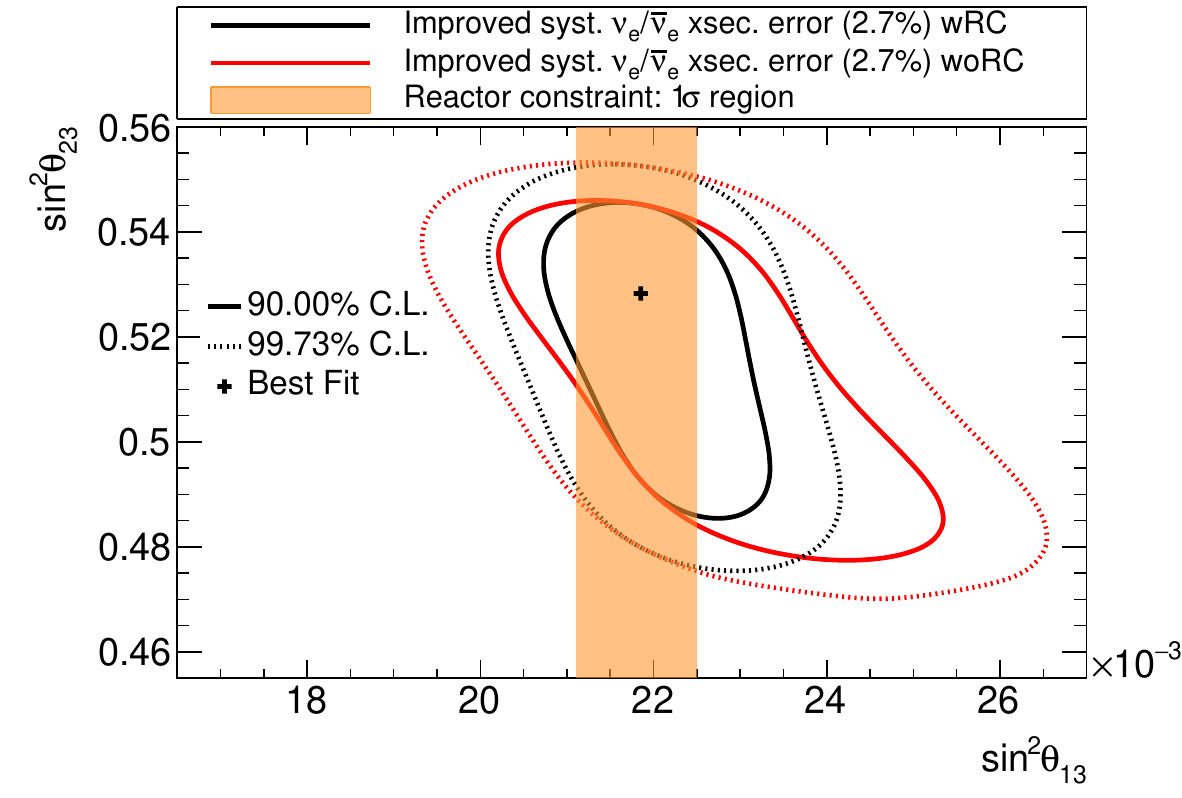}
    \includegraphics[width=0.5\textwidth]{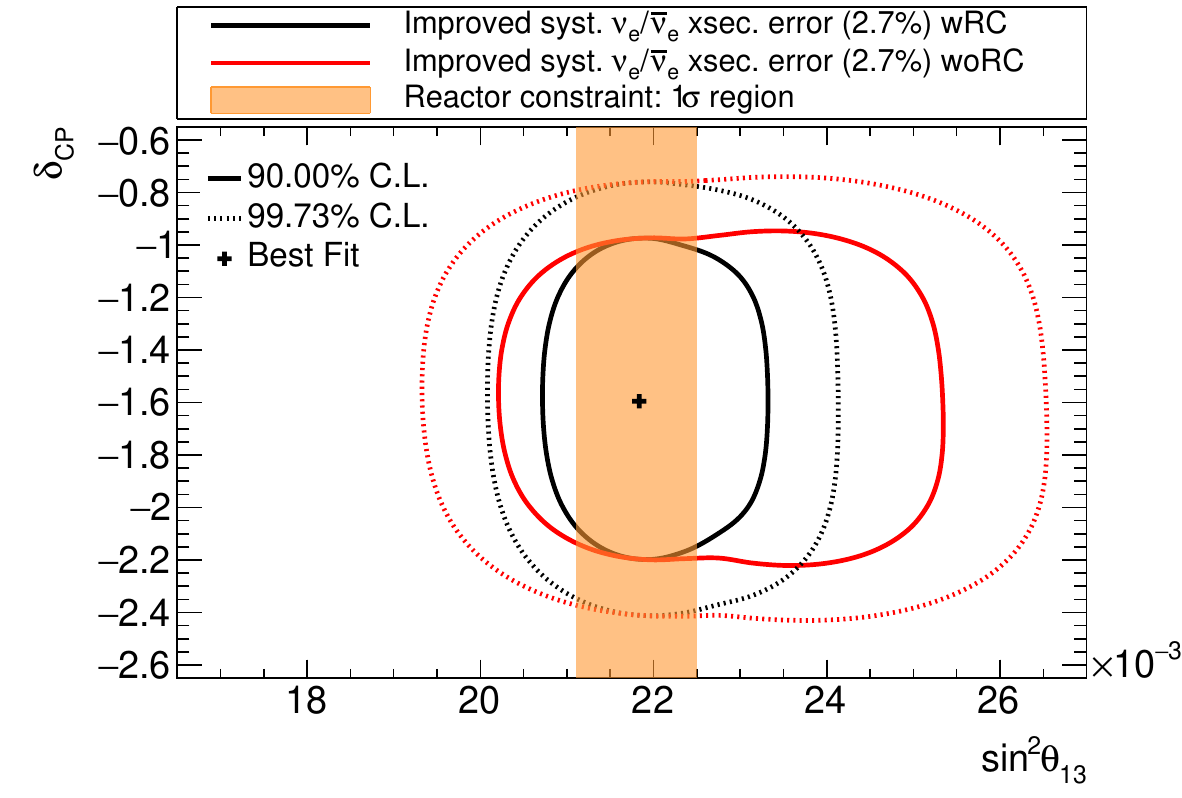}
    \caption{Confidence level contours: $\sin^2\theta_{23}$ vs.\ $\sin^2\theta_{13}$ (top) and $\delta_{CP}$ vs.\ $\sin^2\theta_{13}$ (bottom) after 10 years of data-taking, with the ``Improved syst.'' error model, with (wRC) and without (woRC) external constraint from reactor \(\theta_{13}\) measurements. }
    \label{fig:wrc_worc}
\end{figure}
\begin{table*}[!b]
\begin{tabular}{c c c c c c}
\hline\hline
    Parameter & ~ $\delta_{CP}=0^\circ$ ~ & ~ $\delta_{CP}=-90^\circ$ ~ & ~ $\sin^2\theta_{23}=0.528$ ~ & ~ $\Delta m^2_{32}=2.509\times10^{-3}$ ~ & ~ $\sin^2\theta_{13}=0.0218$ ~ \\
    \& true value & & & & eV$^2$/c$^4$ & (with RC) \\
    \hline
    Statistics only & $5.2^\circ$ & $18.5^\circ$ & $0.0103$ & $7.30\times10^{-6}$& $4.73\times10^{-4}$ \\
     &  &  & $1.95\%$ & $0.29\%$ & $2.17\%$ \\
     \hline
    Improv. systematics & $6.3^\circ$ & $20.2^\circ$ & $0.0134$ & $8.69\times10^{-6}$ & $5.39\times10^{-4}$ \\
     &  &  & $2.54\%$ & $0.35\%$ & $2.47\%$ \\
     \hline
    T2K systematics in~\cite{T2K:2023smv}  & $8.3^\circ$ & $23.9^\circ$ & $0.0199$ & $11.62\times10^{-6}$ & $6.04\times10^{-4}$ \\
     &  &  & $3.77\%$ & $0.46\%$ & $2.77\%$ \\
    \hline \hline
\end{tabular}
\caption{Summary of the 1\(\sigma\) expected resolution of the oscillation parameters after 10 years of data taking. The numbers in percentage are relative errors.}
\label{tab:par_resolutions}
\end{table*}

Finally, the ultimate Hyper-Kamiokande precision on the $\sin^2 \theta_{13}$ parameter is not competitive with measurements from reactor experiments, as shown in Fig.~\ref{fig:s13_wrc_worc}. The $\sin^2 \theta_{13}$ measurement at long baseline experiments has a degeneracy with the $\sin^2\theta_{23}$ parameter, with the two $\theta_{23}$ octants corresponding to the two lobes of the likelihood shown in Fig.~\ref{fig:s13_wrc_worc}. Such degeneracy is also visible in the 2D contours shown in Fig.~\ref{fig:wrc_worc} (top). The reactor constraints solve this degeneracy, allowing Hyper-Kamiokande data to further improve the precision on $\sin^2 \theta_{13}$ by about 15\%.
The 2D contours in $\delta_{CP}$ versus $\sin^2 \theta_{13}$ are shown in Fig.~\ref{fig:wrc_worc} (bottom), where it can be seen that the ultimate Hyper-Kamiokande sensitivity to $\delta_{CP}$ is roughly the same with and without an external constraint from reactor measurements.
Hyper-Kamiokande, indeed, will collect enough statistics in both neutrino and antineutrino modes to probe the possible existence of CP violation independently from the reactor \(\theta_{13}\) measurements.

Table~\ref{tab:par_resolutions} summarizes the expected ultimate precision of the oscillation parameter measurements achievable by the Hyper-Kamiokande experiment.

\section{Conclusion}
\label{sec:conclusions}
This paper describes the analysis to estimate the sensitivity of the Hyper-Kamiokande experiment to long-baseline neutrino oscillation parameters using accelerator (anti)neutrinos. Results are presented for the CPV discovery sensitivity and precision measurements of the oscillation parameters $\delta_{CP}$, $\sin^2\theta_{23}$, $\Delta m^2_{32}$ and $\sin^2\theta_{13}$. This work is based on the T2K analysis in Ref.~\cite{T2K:2023smv}, with tuning applied to the neutrino flux prediction and the far detector simulation to match the Hyper-Kamiokande design~\cite{Hyper-Kamiokande:2018ofw}.  Different assumptions for the systematic uncertainties are compared, starting with the T2K uncertainties from Ref.~\cite{T2K:2023smv} and applying further reductions based on the Hyper-Kamiokande expected statistics and the upgraded ND280 and IWCD capabilities.

With the assumed Hyper-Kamiokande running plan, a 5$\sigma$ CPV discovery is possible in less than three years in the case of maximal CPV and known MO. In the absence of external constraints on the MO, considering the MO sensitivity of the Hyper-Kamiokande measurement using atmospheric neutrinos, the time for a CPV discovery could be estimated to be around six years. We defer a detailed joint Hyper-Kamiokande beam-atmospheric neutrino analysis to a future publication.

Using the nominal final exposure of $27 \times 10^{21}$ protons on target, corresponding to 10 years, with a ratio of 1:3 in neutrino to antineutrino beam mode, we expect to select approximately 10000 charged current, quasi-elastic-like, muon neutrino events, and a similar number of muon antineutrino events.  In the electron  (anti)neutrino appearance channels, we expect approximately 2000 charged current, quasi-elastic-like electron neutrino events and 800 electron antineutrino events, assuming $\delta_{CP} = -1.601$.  These large event samples will allow Hyper-Kamiokande to exclude CP conservation at the $5\sigma$ significance level for over 60\% of the possible true values of $\delta_{CP}$.  Depending on the value of $\delta_{CP}$, Hyper-Kamiokande can measure the $\delta_{CP}$ parameter to a precision of about $6\degree$ (in the case of CP conservation) or $20\degree$ (in the case of maximal CPV).
The wrong $\sin^2\theta_{23}$ octant can be excluded at a significance above $5\sigma$ for $\sin^2\theta_{23} < 0.45$ and $\sin^2\theta_{23} > 0.57$.  The value of $\sin^2\theta_{23}$ can be measured with a precision of around 3\% in the most challenging region of maximal disappearance and better than 0.5\% otherwise. The expected precision on the measurement of $\Delta m^2_{32}$ is better than 0.5\%.

With the assumed running ratio of 1:3 for neutrino to antineutrino beam mode operation, the Hyper-Kamiokande ultimate $\delta_{CP}$ resolution is mainly independent of the constraint on $\sin^2\theta_{13}$ from external reactor measurements. Still, this reactor constraint reduces the degeneracy between $\sin^2\theta_{13}$ and $\sin^2\theta_{23}$. When the reactor constraint on $\sin^2\theta_{13}$ is applied and thus the degeneracy resolved, Hyper-Kamiokande will be able to slightly improve the precision on the measurement of $\sin^2\theta_{13}$ with respect to the reactor measurement.
\section*{Acknowledgements}

The Hyper-Kamiokande Collaboration would like to thank
the Japanese Ministry of Education, Culture, Sports, Science and Technology (MEXT), the University of Tokyo, Japan Society for the Promotion of Science (JSPS), the Kamioka Mining and Smelting Company, Japan; 
% Armenia
the Minister of Education, Science, Culture and Sports, grant 21T-1C333, Armenia;
% Australia
%, Australia;
% Brazil
CNPq and CAPES, Brazil;
% Canada
%, Canada;
% Czech Republic
Ministry of Education Youth and Sports of the Czech Republic 
(FORTE {CZ.02.01.01/00/22\_008/0004632}); %, Czech Republic;
% France
CEA and CNRS/IN2P3, France;
% Germany
%, Germany;
% Greece
%, Greece;
% India
%, India;
% Italy
the INFN, Italy;
% % Korea
the National Science Foundation, the Korea National Research Foundation (No. NRF 2022R1A3B1078756), Korea;
% MExico
CONAHCyT for supporting the national projects CBF2023-2024-427, and CF-2023-G-643., Mexico;
% Morocco
%, Morocco;
% Poland
Ministry of Science and Higher Education (2022/WK/15) and the National Science Centre (UMO-2022/46/E/ST2/00336), Poland;
% Russia
% Spain
MICIU, Spain and NextGenerationEU/PRTR, EU, Spain;
% Sweden
%, Sweden;
% Switzeland
ETHZ, SERI and SNSF, Switzerland;
% UK
STFC and UKRI, UK;
% Ukraine
%, Ukraine;
% US
We also thank CERN for use of the Neutrino Platform.
% Computing
% Individuals
In addition, the participation of individual researchers and institutions has been further supported by funds from  
% Czechia Republic
Charles University (Grant PRIMUS 23/SCI/025 and Research Center UNCE/24/SCI/016), Czech Republic;
%Europe
 H2020-MSCA-RISE-2019 SK2HK no 872549 and H2020 Grant No. RISE-GA822070-JENNIFER2 2020, Europe;
 % Japan
 JSPS KAKENHI Grant JP24K17065; 
 % Korea
National Research Foundation of Korea (NRF-2022R1A5A1030700, NRF-2022R1A3B1078756) funded by the Ministry of Science, Information and Communication Technology (ICT) and the Ministry of Education (RS-2024-00442775), Korea;
% MExico
Associate Dean of Research and Scientific Graduate Studies, Dr. Daniel A. Jacobo-Velázquez, for his support and also to CONAHCyT for the postgraduate scholarship 835328; Rector of the CUCEI, Dr. Marcos Pérez, for his financial and logistical support and CONAHCyT for the information technologies doctoral scholarship 792151, Mexico;
% Poland
Ministry of Science and Higher Education, Republic of Poland, "International co-financed projects", grant number 5316/H2020/2022/2023/2, Poland;
% Spain
LSC funds from MICIU, DGA and UZ, Spanish Ministry of Science and Innovation PID2022-136297NB-I00 /AEI/10.13039/501100011033/ FEDER, UE; CERCA program of the Generalitat de Catalunya; Plan de Doctorados Industriales of the Research and Universities Department of the Catalan Government (2022 DI 011); MICIIN; European Union NextGenerationEU(PRTR-C17.I1); Generalitat de Catalunya;
Spanish Ministry of Innovation and Science under grants MCINN-23-PID2022-139198NB-I00 and PID2021-124050NB-C31, Spain;
% Switzerland
SNF 20FL20-216674, SNF 200021L-231581 and SNF PCEFP2-203261, Switzerland;
% UK
the Leverhulme Trust Research Fellowship Scheme and the 
UKRI Future Leaders Fellowship grant number MR/S034102/1, UK.

%The Hyper-Kamiokande Collaboration would like to thank the T2K Collaboration for providing the models and simulated samples, as well as analysis software originally developed within T2K, which were used as a basis for this analysis.  
%The Hyper-Kamiokande Collaboration would like to thank the T2K Collaboration for providing the models and simulated samples which were used as a basis for this analysis.  
The Hyper-Kamiokande Collaboration would like to thank the T2K Collaboration for providing the models and simulated samples used as a basis in this paper.  

\bibliographystyle{spphys}       % APS-like style for physics
\typeout{}
\bibliography{biblio}
%\author{HK collaboration}

%\input{authors}

\end{document}